\PassOptionsToPackage{pdfpagelabels=false}{hyperref}
\documentclass[useAMS,usenatbib]{mnras}
\pdfoutput=1

\usepackage{newtxtext,newtxmath}
\usepackage{newtxtext,newtxmath}

\usepackage[T1]{fontenc}
\usepackage{ae,aecompl}

\usepackage{amsmath}	
\usepackage{amssymb}	
\usepackage{siunitx}
\usepackage[utf8]{inputenc}
\usepackage{xcolor}
\usepackage{times}
\usepackage{color}
\usepackage[pdftex]{graphicx}
\usepackage{natbib}
\usepackage{comment}
\usepackage[normalem]{ulem}

\def \MSUN{{\rm M}_{\odot}}

\graphicspath{ {./images/} }
\DeclareSIUnit \parsec {pc}
\DeclareSIUnit\year{yr}
\DeclareSIUnit \mega {M}

\title[The IRX--$\beta$ dust attenuation relation in TNG50]{A redshift--dependent IRX--$\beta$ dust attenuation relation for TNG50 galaxies}

\author[S. Schulz et al.]{Sebastian Schulz$^{1}$\thanks{E-mail: sebastian-schulz713@web.de},
Gerg\"o Popping$^{2,1}$\thanks{E-mail: gpopping@eso.org},
Annalisa Pillepich$^{1}$,
Dylan Nelson$^{3}$,\newauthor
Mark Vogelsberger$^{4}$,
Federico Marinacci$^{5}$,
Lars Hernquist$^{6}$
\\
\\
$^{1}$Max-Planck-Institut für Astronomie, Königstuhl 17, D-69117 Heidelberg, Germany\\
$^{2}$European Southern Observatory, Karl-Schwarzschild-Str. 2, D-85748, Garching, Germany\\
$^{3}$Max-Planck-Institut f\"u Astrophysik, Karl-Schwarzschild-Str. 1, D-85748 Garching, Germany\\
$^{4}$Kavli Institute for Astrophysics and Space Research, Department of Physics, MIT, Cambridge, MA, 02139, USA\\
$^{5}$Department of Physics and Astronomy, University of Bologna, via Gobetti 93/2, 40129 Bologna, Italy\\
$^{6}$Harvard-Smithsonian Center for Astrophysics, 60 Garden Street, Cambridge, MA, 02138, USA}

\date{Accepted XXX. Received YYY; in original form ZZZ}

\pubyear{2019}

\begin{document}
\label{firstpage}
\pagerange{\pageref{firstpage}--\pageref{lastpage}}
\maketitle

\begin{abstract}
We study the relation between the UV--slope, $\beta$, and the ratio between the infrared-- and UV--luminosities (IRX) of galaxies from TNG50, the latest installment of the IllustrisTNG galaxy formation simulations. We select 7280 star--forming main--sequence (SFMS) galaxies with stellar mass $\geq10^9\MSUN$ at redshifts $0 \leq z \leq 4$ and perform radiative transfer with {\sc skirt} to model effects of interstellar medium dust on the emitted stellar light. Assuming a Milky Way (MW) dust type and a dust--to--metal ratio of 0.3, we find that TNG50 SFMS galaxies generally agree with observationally--derived IRX--$\beta$ relations at $z \lesssim 1$. However, we find a redshift--dependent systematic offset with respect to empirically--derived local relations, with the TNG50 IRX--$\beta$ relation  shifting towards lower $\beta$ and steepening at higher redshifts. This is partially driven by variations in the dust--uncorrected UV--slope of galaxies, due to different star--formation histories of galaxies selected at different cosmic epochs; we suggest the remainder of the effect is caused by differences in the effective dust attenuation curves (EDACs) of galaxies as a function of redshift. We find a typical galaxy--to--galaxy variation of 0.3 dex in IRX at fixed $\beta$, correlated with intrinsic galaxy properties: galaxies with higher star--formation rates, star--formation efficiencies, gas metallicities and stellar masses exhibit larger IRX values. 
We demonstrate a degeneracy between stellar age, dust geometry and dust composition: $z=4$ galaxies with a Small Magellanic Cloud dust type follow the same IRX--$\beta$ relation as low--redshift galaxies with MW dust. 
We provide a redshift--dependent fitting function for the IRX--$\beta$ relation for MW dust based on our models.

\end{abstract}




\section{Introduction} \label{Introduction}

Determining the cosmic star formation rate density (SFRD) and its evolution with redshift is of critical importance for our understanding of galaxy formation and evolution. First measurements have shown that the cosmic SFRD has been steadily decreasing since redshift $z\approx$ 1 (\citealt{Lilly}, \citealt{Madau}). More recent observations have shown that the cosmic SFRD reaches a peak at $z \approx 2$, declining again as we go towards higher redshifts \citep[see the review by][]{Madau2}. Measurements of star formation rates (SFRs) at $z > 3$ are often based on the ultraviolet (UV) emission as a tracer for the abundance of young stars in a galaxy, and therefore as a tracer for SFR (\citealt{Kennicutt}). The attenuation of stellar light by dust in the interstellar medium (ISM) is a major hindrance at these wavelengths.  UV-emission is being scattered and absorbed, changing the measured shape and luminosity of the UV spectrum. The absorbed UV--light is being re--emitted in the infrared (IR) regime,
and ideally we can account for the absorbed emission by measuring both the UV spectrum and the IR spectrum of a galaxy (e.g. \citealt{Reddy}). At high redshifts $z > 3$, reliable IR data is often missing, which is why alternatives for accounting for UV dust attenuation have been explored.

For local UV--bright starburst galaxies, an empirically derived tight relation between the rest--frame UV continuum slope $\beta$ (where we assume a power law: $f_\lambda \propto \lambda^{\beta}$) and the ratio between IR and UV luminosity, called the infrared excess (IRX) has been found (\citealt{Meurer}). This IRX-$\beta$ relation provides a connection between the shape of the UV continuum (the "color" of the galaxy) and the amount of dust attenuation. If this relation proved to hold for various galaxy types at different redshifts, it could be used as a reliable tool to account for dust attenuation where only UV data is available and IR data is missing (e.g. as in \citealt{Bouwens}, \citealt{Oesch}, \citealt{McLeod}). 

Recent studies have investigated the IRX--$\beta$ relation of various galaxy types and at diverse redshifts. Some adjustments to the Meurer relation have been proposed to correct for biases in aperture size and sample selection (\citealt{Overzier_2010}, \citealt{Takeuchi_2012}, \citealt{Casey_2014}). These corrected relations seem to agree qualitatively at $z = 0$ for galaxies on the star forming main sequence (SFMS), and also agree with observations of main sequence galaxies at $ z< 4$ (e.g. \citealt{Heinis2013}, \citealt{Bouwens2016}, \citealt{AM2016}, \citealt{Bourne2017}, \citealt{Fudamoto2017}, \citealt{Koprowski2018}, \citealt{McLure}, \citealt{Reddy}, \citealt{Fudamoto2019}, Bouwens et al. in prep.). However, allowing for a more complete sample of galaxies including IR--bright starbursts and submillimeter galaxies (SMGs) and going towards higher redshifts  increases the scatter in the IRX--$\beta$ plane (\citealt{Kong}, \citealt{Seibert_2005}, \citealt{Takeuchi_2010}, \citealt{Oteo2013}, \citealt{Casey_2014}). 

Various groups have ventured to quantify this scatter and find the physical drivers of it. For instance, the broad distribution of stellar population ages is presumed to affect the resulting IRX--$\beta$ of a galaxy, increasing the scatter and leading to a systematic shift along the $\beta$--axis as the median stellar population age varies across galaxies (with $\beta$ increasing with stellar population age, \citealt{Kong}, \citealt{Boquien_2009}).  Additionally, scatter is expected from variations in the effective dust attenuation curves (EDAC) of galaxies (see review by \citealt{salim2020dust} and recent findings by \citealt{Salim_2019}). EDACs are an extension to the usual extinction curves known from homogeneous dust screens, with each galaxy's EDAC arising from a combination of its dust column density, its dust grain composition, and the geometric relationship between its stellar and dust distribution. The EDACs influence the shape of the attenuated galaxy spectra and hence also the shape of the respective IRX--$\beta$ relations.

Measurements of dusty star forming galaxies (DSFGs) and ultra luminous infrared galaxies (ULIRGs), which are characterized by a central star forming region the size of a few hundred pc, and 90\% of their energy output being in the infrared regime, have shown that they have significantly higher IRX than usual galaxies, when compared to normal local galaxies with similar $\beta$ (\citealt{Casey_2014}, \citealt{Goldader}). A proposed explanation for this is that these highly star forming galaxies have very complex geometries regarding their dust distribution, leading to parts of the UV emission being only weakly attenuated, with other parts being heavily obscured by dust (\citealt{Calzetti2001}, \citealt{Casey_2014}). This leads to a UV continuum dominated by young stars not obscured by dust, causing low values of $\beta$, while the IR continuum is being dominated by regions with heavy dust obscuration, causing high IRX (see also the theoretical explanations in \citealt{Popping} and \citealt{Narayanan}). 

Some galaxies lie significantly below and to the right of the original Meurer relation. It has been proposed that the grain type- and size-distribution of the ISM dust affect the exact location in the IRX-$\beta$ plane, by influencing the far UV (FUV) extinction curve (\citealt{Pettini}, \citealt{Witt}). For example, dust as it is present in the small and large magellanic clouds (SMC and LMC) has a far higher FUV extinction than Milky Way dust, which leads to a stronger dependence of $\beta$ on the dust opacity. Ultimately, this translates to a less steep IRX--$\beta$ relation for galaxies with strongly FUV--extincting dust types. 

In short, variations in dust properties, dust geometry, as well as stellar population age all affect the location of galaxies in the IRX-$\beta$ plane, and this leads to an apparent scatter which is present at all redshifts. Other drivers for the observed scatter, such as dust temperature \citep{Narayanan, Faisst}, or variations in intrinsic $\beta_0$ (the slope of the UV continuum without any dust attenuation, \citealt{Boquien_2012}) have also been suggested, but a definite answer as to which are the main mechanisms behind the observed scatter is yet to be found. 
The thruth may lie in a combination of all proposed explanations and in the fact that normal star forming galaxies typically consist of many star forming regions disconnected from each other, each with their own stellar population ages, dust types and dust geometries.
  
Apart from observational investigations, analytical models (e.g., \citealt{Popping}), employing two component dust models and stellar population synthesis have confirmed the previously mentioned ideas of causes for scatter in the IRX--$\beta$ plane. However, even though analytical modeling gives us valuable insights into the effects several physical properties have on the IRX--$\beta$ scatter in a controlled environment, it only provides a simplified view of the subject, and struggles to model the complex interplay of the many variables at hand. 

Numerical simulations have succeeded to account for the complex dynamics of galaxies and the interplay between various physical quantities by coupling idealized (\citealt{Safarzadeh}) or zoom-in simulations (\citealt{Narayanan}, \citealt{Ma}) with radiative transfer calculations. These works confirmed a tight IRX--$\beta$ relation for Milky Way type galaxies even at higher redshifts, as well as higher dust temperatures in high--redshift galaxies partly due to increased star formation rates. The simulations also suggest that certain galaxy types such as starbursts, DSFGs and ULIRGs lie above the Meurer relation, due to complex dust geometries that lead to spatially disconnected UV-- and IR--emission. In these cases the attenuated UV--emission is often spatially very extended and dim. 

When it comes to numerically simulating galaxies, there have been three main types of simulations: (i) idealized simulations, that simulate single isolated galaxies or galaxy mergers without any cosmological context, (ii) zoom--in simulations, that trace the evolution of small samples of galaxies in a cosmological context through time, with very high resolution ($M_{\rm baryons}\lesssim 10 ^4 $M$_\odot$), and initial conditions taken from large volume dark matter only simulations, or (iii) full physics large volume cosmological simulations, that start from primordial initial conditions and trace the evolution of a large sets of galaxies in a large cosmic volume through time, with intermediate resolutions ($M_{\rm baryons} \approx 10^6 $M$_\odot$). The reason that other groups have focused on idealized or zoom--in simulations to study the IRX--$\beta$ relation is simple: complex dust geometries promise to be an explanation for the IRX--$\beta$ scatter across galaxies, and these geometries had to be resolved well enough in simulations. So far such resolutions have only been achievable with these two simulation types. The downside of using zoom--in simulations is that in order to achieve these high resolutions, they are restricted to small galaxy samples, not necessarily representative of the overall galaxy population. On the other hand, typical large scale simulations have not been an ideal option for IRX--$\beta$ investigations, as their numerical resolution was too low and did not properly capture the complex distribution of stars and the ISM. 

This has changed now, with the new TNG50 simulation (\citealt{Nelson50}, \citealt{Pillepich50}). It is the highest resolution variant of the IllustrisTNG simulation suite (\citealt{Marinacci}, \citealt{Naiman1}, \citealt{Nelson}, \citealt{Pillepich}, \citealt{Springel}), and has a mass resolution of $M_{\rm baryons} \sim 8.5 \times 10^{4} \MSUN$ in a cosmological volume of $\sim$ 50 cMpc on a side, bridging the gap between zoom--in and large scale cosmological simulations, making it ideal to study a representative sample of galaxies over cosmic time at sufficiently good mass resolution. It is especially well suited to this work, because it combines a high resolution required to accurately model the dust distribution of each galaxy with a sufficiently large sample size of galaxies which is needed to perform a sensible analysis of the problem at hand. 

In this work, we will try to quantify the scatter in the IRX--$\beta$ dust attenuation relation. 
We will investigate most of the star forming galaxies in the cosmological volume of TNG50 between the current epoch and $z=4$. We employ the radiative transfer code {\sc skirt} (\citealt{Baes_2011}, \citealt{Camps}) to model the attenuation of stellar light by dust and obtain full spectra with high detail in UV and IR. We will calculate IRX and $\beta$ for all of these galaxies and explore how various physical and observable quantities of those galaxies, which are directly taken or derived from the TNG50 simulation data, affect the location of those galaxies in the IRX--$\beta$ plane. We will also study the effect different dust properties have on the location of galaxies in the IRX--$\beta$ plane.

This paper is organized as follows. In Section \ref{Methods} we describe the numerical framework by introducing the IllustrisTNG simulation suite and by describing the galaxy selection, as well as the {\sc skirt} radiative transfer code. In such section, we also explain how the IRX and $\beta$ are derived. In Section \ref{Results}, we show our main results, quantifying the scatter and locus of the IRX-$\beta$ relation. Section \ref{Discussion} contains a discussion and possible interpretations of our findings. It is followed by an outlook on ideas for future work and finally a summary of our main results in Section \ref{Summary}. 
Appendix \ref{Appendix} contains various checks on the dependence of the fiducial radiative transfer model on a number of implementation choices.

\section{Methods} \label{Methods}

\subsection{The TNG50 Simulation} \label{TNG50}

The IllustrisTNG Project \citep[TNG hereafter:][]{Nelson, Marinacci, Springel, Naiman1, Pillepich} is a suite of magneto--hydrodynamical cosmological simulations for the formation of galaxies employing the moving mesh {\sc arepo} \citep{SpringelA} code. In this work, we use the output of its highest--resolution implementation: the TNG50 simulation \citep{Nelson50, Pillepich50}. 

The TNG project is based on an updated version of the former Illustris galaxy formation model (\citealt{Vogelsberger}, \citealt{Vogelsberger2014}, \citealt{Torrey}) described in \citealt{Weinberger} and \citealt{PillepichTNG}. The updates it received in comparison to the original Illustris model include the addition of magneto--hydrodynamics, a new model of Active Galactic Nuclei (AGN) feedback for low accretion rates (\citealt{Weinberger}), and modifications to the galactic winds, stellar evolution and chemical enrichment schemes (\citealt{PillepichTNG}). The simulations trace the evolution in time of dark matter particles, stellar particles and stellar winds, gas cells and black holes, starting from initial conditions at $z = 127$, ending at $z = 0$. Identification of haloes and subhaloes was done with the Friends--of--Friends (FoF, \citealt{Davis}) and {\sc subfind} (\citealt{SUBFIND1}, \citealt{SUBFIND2}) algorithms. The subhaloes within each FoF halo contain all resolution elements that are gravitationally bound to them. Within the TNG50 framework, we consider those subhaloes to be galaxies which have non-vanishing stellar mass and a total dark matter mass fraction of at least 20\% (see \citealt{NelsonTNG}, Section 5.2; or \citealt{Pillepich50}, Section 3.1). The IllustrisTNG model roughly matches the observed cosmic SFRD for $0 < z < 8$, the galaxy mass function at $z = 0$, the stellar--to--halo mass relation at $z =0$, the halo gas fraction at $z = 0$ and galaxy sizes at $z = 0$ (see \citealt{Pillepich_2017}). IllustrisTNG uses cosmological parameters consistent with the values recently measured by the Planck collaboration (\citealt{Planck_2015}).

There are three different flagship realizations of the IllustrisTNG suite, namely TNG300, TNG100 and TNG50 - named after their simulation box sizes at $z=0$ of $50$, $100$ and 300 comoving Mpc, respectively. They provide different mass resolutions and were designed for different purposes: TNG300 simulates the largest cosmological volume of them all, and thus provides insights on the large scale physics with a large number of simulated galaxies as well as rare objects, like massive galaxy clusters. TNG50, with its smaller volume delivers relatively less statistics, but provides a numerical resolution that is comparable to those of zoom-in simulations, making it possible to study detailed structures of galaxies. TNG100 lies in between the two. All variations also have lower--resolution variants (so that resolution effects can be quantified), and dark matter--only variants. In this work, we use data of the highest resolution version TNG50 (\citealt{Nelson50}, \citealt{Pillepich50}, see also Table \ref{tab:numerics}).

TNG50 follows the evolution of $2 \times 2160^{3}$ resolution elements inside a cube measuring 35 cMpc/h on each side, which correpsonds to a volume of 51.7$^{3}$ (cMpc)$^{3}$. This translates to a target mass resolution of  $0.85 \times 10{^5} \MSUN$ for the baryonic resolution elements (gas cells and stellar particles) and to $4.5 \times 10{^5} \MSUN$ for the dark matter resolution elements. Subscale physical mechanisms (below 70-150 parsecs within the star forming regions of galaxies, see Figure 1 of \citealt{Pillepich50}) are included via subgrid modeling, this being the case for example for star formation, supernovae and AGN feedback (\citealt{Hernquist}, \citealt{PillepichTNG}, \citealt{Weinberger}). In this simulation, galaxies with stellar mass $M_* = 10^{9}\MSUN$ and above will contain at least about 10,000 stellar particles, so their resolution is in between those of typical cosmological simulations and those of zoom--in simulations.  
We will make use of a number of TNG50 data output, including various physical quantities of stellar particles and gas cells as well as integrated properties of the halos and subhaloes identified by the {\sc subfind} algorithm. We will later use these data to identify drivers of scatter in the IRX--$\beta$ relation.
  
\begin{figure*}
    \centering
    \includegraphics[width = 0.85\hsize]{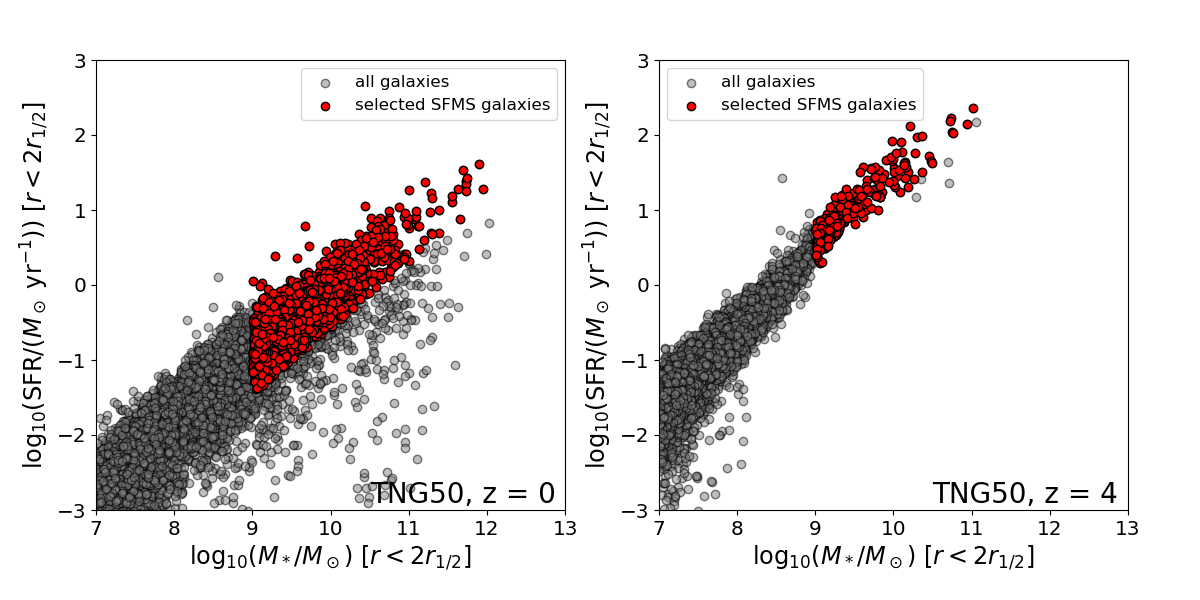}
    \caption{The selection of TNG50 galaxies used in this work. Here we show the TNG50 galaxy population at the edges of the considered redshift range, $z=0$ (left) and $z = 4$ (right), according to the galaxies' star formation activity and stellar mass (where SFR and stellar mass are taken inside twice the 3D stellar half mass radius $r_{1/2}$). Gray circles denote all TNG50 galaxies, red filled circles represent those we select for this work: star-forming and well resolved i.e. with $M_* > 10^{9} \MSUN$). Notice that as we increase redshift, fewer galaxies are found in the selected mass range (see also Table \ref{tab:tab1}).}
    \label{fig:fig1}
\end{figure*}  
  
\subsection{TNG50 Galaxy Selection} 
\label{Galaxy Selection}
In this work, we focus on star--forming galaxies, meaning those that lie on the SFMS and above. The classification of SFMS galaxies is done as in \citealt{Pillepich50}, Section 4. 

In short, we stack all TNG50 galaxies into 0.2 dex stellar mass bins in the range $M_* = 10^8 \MSUN \text{ to } 10^{10.2} \MSUN$. Then, we calculate the median specific star formation rate (sSFR) inside twice the 3D stellar half mass radius $r_{1/2}$ in each bin. The sSFR of a galaxy is defined as the instantaneous star formation rate (SFR) inside $2 r_{1/2}$ divided by the stellar mass inside $2 r_{1/2}$. This sSFR median defines the SFMS ridge, and green--valley and quenched galaxies are "rejected" as those that have a logarithmic distance of $-1 < \Delta \log_{10} \text{sSFR} < -0.5$ or $\Delta \log_{10} \text{sSFR} < -1$, respectively. The process is then repeated, without considering green valley and quenched galaxies, until the median converges. A power law is then fitted on the sSFR median to extrapolate the SFMS to higher masses $M_* > 10^{10.2}\MSUN$. The galaxies with $\Delta \log_{10} \text{sSFR} > -0.5$ are then the star--forming galaxies of the SFMS. 

We make no distinction between central and satellite galaxies. We do not inspect low--SFR galaxies, because their dust content is expected to be comparably low, making them less prone to dust attenuation effects. We note that this transition is gradual, as for example green valley galaxies' stellar light can be notably attenuated by dust, however we stick to the well--defined sample of SFMS galaxies in our investigations. We select star--forming galaxies across different redshifts ($z$ = 0, 0.5, 1, 2, 3, 4), so as to investigate a possible redshift--dependent evolution in the IRX--$\beta$ relation.  
We limit ourselves to redshifts $z \leq 4$, because at higher redshifts the effect of the cosmic microwave background (CMB) on the infrared emission of galaxies, which is not accounted for in the radiative transfer modeling employed in this work, becomes non-negligible. 

We investigate galaxies that are made of at least 10,000 stellar particles to ensure that they are well enough resolved to reproduce complex dust geometries. This translates to a minimum stellar mass of $M_* \gtrapprox 10^9 \MSUN$ for TNG50. Notice that as we increase redshift, the median stellar mass of galaxies is decreasing, so the higher we go in redshift, the smaller the sample size will be given our selection cut of $M_* > 10^{9}\MSUN$ (see Table \ref{tab:tab1}). In total, we end up with 7280 well resolved SFMS galaxies in the redshift range $0 \leq z \leq 4$.

  \begin{table}
  \centering
  \resizebox{\columnwidth}{!}{%
	\begin{tabular}{ccc} 
		\hline
		Redshift & \# TNG50 galaxies: & \# TNG50 galaxies:  \\
		$z$ & $M_* > 10^9 \MSUN$ & SFMS and $M_* > 10^9 \MSUN$)  \\
		\hline
		0.0 & 2524 & 1790 \\
		0.5 & 2366& 1767 \\
        1.0     & 2090 & 1710\\
        2.0     & 1272 & 1149\\
        3.0     & 652 & 620\\
        4.0     & 249 & 244\\
		\hline
	\end{tabular}
	}
	\caption{ A summary of the number of TNG50 galaxies used in this work. From left to right: redshift, total number of galaxies with mass $M_* > 10^9 \MSUN$, number of selected SFMS galaxies given our selection criteria at the given redshift. At high redshifts, we have in total fewer galaxies, but a higher percentage of them belongs to the SFMS.}
	\label{tab:tab1}
\end{table}

\subsection{Obtaining TNG50 Galaxy Spectra with {\sc skirt}}\label{Obtaining Spectra}

The IllustrisTNG simulation suite does not track radiative processes. Hence, in order to accurately model attenuation of TNG50 galaxy spectra by dust in the ISM, we employ radiative transfer calculations in post--processing. We make use of the publicly available radiative transfer code {\sc skirt}\footnote{{\sc skirt} homepage: \url{http://www.SKIRT.ugent.be}} (\citealt{Baes_2011}, \citealt{Camps}), which performs a Monte--Carlo tracing of the paths of photons emitted by a stellar distribution as they are scattered and absorbed by a dust distribution. These photons are captured by a virtual observing instrument which is capable of returning the total spectral energy distribution (SED) of the captured photons, as well as a spatially resolved spectrum (a data cube containing a 2D image of the observed fluxes at each specified wavelength). The stellar distribution used in our {\sc skirt} calculations is predicted by the TNG model and thus directly imported from the TNG50 simulation output. We note that ISM dust is not tracked in the IllustrisTNG simulations. We will therefore assume that the simulations' metal distribution is a tracer for the ISM dust distribution, following the approach of similar studies (e.g. \citealt{CampsSKIRT}, \citealt{Trayford}, \citealt{Ma}, \citealt{Gomez}, \citealt{Vogelsberger2019}, see Section 
\ref{Dust Modelling} for more details).

\subsubsection{Stellar Sources}\label{Stellar Sources}
Each stellar particle in TNG50 is a point--like coeval stellar population. For the treatment with {\sc skirt}, we define an adaptive smoothing scale to each stellar particle equal to the 3D distance to the $N$--th nearest neighbouring stellar particle, which has been described in \citealt{Torrey}. This is to allow {\sc skirt} to calculate internally a smoothed photon source distribution function. \citealt{Torrey} use $N=16$, however we orient ourselves on \citealt{Gomez}, who have chosen $N = 32$ and have also found that varying $N$ in the range of 4 to 64 has no particular effect at least on morphological measurements. We tested if this is the case for IRX--$\beta$ measurements, too, and found no noticeable difference between choosing $N$ = 16, 32 or 64.

All smoothing scale calculations are carried out by the default smoothed particle hydrodynamics (SPH) spline kernel (\citealt{Monaghan}), redefined over the interval $[0,h_{sml}]$ (\citealt{SpringelGA}), where $h_{sml}$ is the smoothing scale described before:
\begin{equation}
    W(r,h_{sml}) = \frac{2^\nu \sigma}{h^n_{sml}}\begin{cases}1-6q^2+6q^3 & \text{if\ \ \ } 0 \leq q \leq \frac{1}{2} \\
    2(1-q)^3 & \text{if\ \ \ } \frac{1}{2} < q \leq 1 \\
    0 & \text{if\ \ \ } q > 1
    \end{cases}
\end{equation}
where $q \equiv r/h_{sml}$, $r$ is the radial distance, $n$ is the number of dimensions, and $\sigma$ is a normalization constant with the values $2/3$, $10/7\pi$, $1/\pi$, in one, two and three dimensions, respectively.

The SEDs of stellar particles older than 10 Myr are modeled with the {\sc galaxev} population synthesis code (\citealt{Bruzual}), which is implemented internally in {\sc skirt}. This code uses simple stellar population (SSP) models, which were computed using Padova 1994 evolutionary tracks and a \citealt{Chabrier} initial mass function (IMF). These models provide the rest--frame luminosity per unit wavelength of an SSP, $L_\lambda(\lambda, t, Z)$ as a function of wavelength $\lambda$, age $t$, and metallicity $Z$ (the luminosity is normalized to solar mass $\MSUN$). The luminosity is provided by an interpolation of a grid which is sampled at 221 unevenly spaced ages between 0 and 20 Gyr, 7 metallicities between $10^{-4}$ and 0.5 and 1221 wavelengths between  $\SI{91}{\angstrom}$ and $\SI{160}{\micro \meter}$.
To generate a stellar spectrum, the {\sc galaxev} code requires as an input the initial mass of the stellar population (neglecting mass loss due to stellar evolution), its metallicity, and stellar age.

Young stellar populations present in starbursting regions are formed and embedded in dense and cold molecular dust, which we call "birth clouds" (\citealt{Charlot_2000}). As the lifetime of these molecular birth clouds is about 10 Myr (\citealt{1980ApJ...238..148B}), we assume that stellar populations younger than that are still surrounded by these clouds. These clouds are not resolved by TNG50, as their sizes are on pc scales. To account for attenuation in birth clouds, all stars younger than 10 Myr will be treated as starbursting regions in this work, and their SEDs are modeled using the {\sc mappings-iii} photoionization code (\citealt{Groves}), which includes emission from HII--regions and their surrounding photodissociation--regions (PDRs), as well as absorption by gas and dust in the birth clouds. The {\sc mappings-iii} models require the following five input parameters for each star forming region: (i) The SFR, assumed to be constant over the last 10 Myr, (ii) the metallicity, (iii) the compactness parameter $C$, (iv) the ISM pressure $P_\text{ISM}$ and finally (v) the PDR covering fraction $f_\text{PDR}$. For every young stellar particle, we assume that the SFR is given by its initial mass divided by 10 Myr (to ensure mass conservation), and use its nominal metallicity inherited from its parent gas cell. For the compactness $C$ and the ISM pressure $P_\text{ISM}$, we use typical values found in literature (\citealt{Groves}) of $\log_{10}C = 5$ and $\log_{10} \frac{P_\text{ISM}/k_B}{\SI{}{\centi \metre^{-3}}\SI{}{\kelvin}} = 5$. The exact values of these two parameters only have a noticeable influence on the far--IR SEDs, we tested what effect a variation of these values has on the result of IRX--$\beta$ (see Appendix \ref{AppMappings}). While it makes a difference whether we include {\sc mappings-iii} or not, variations of $C$ and $P_\text{ISM}$ have only marginal effects on the resulting total IRX--$\beta$ of a galaxy. Following \citealt{Jonsson10} and \citealt{Gomez}, we adopt a value of 0.2 for the covering fraction $f_\text{PDR}$ for our fiducial model. Variation of this value also has no significant qualitative impact on the results (see Appendix \ref{AppMappings}).

\begin{figure*}
    \centering
    \includegraphics[width = 0.85\hsize]{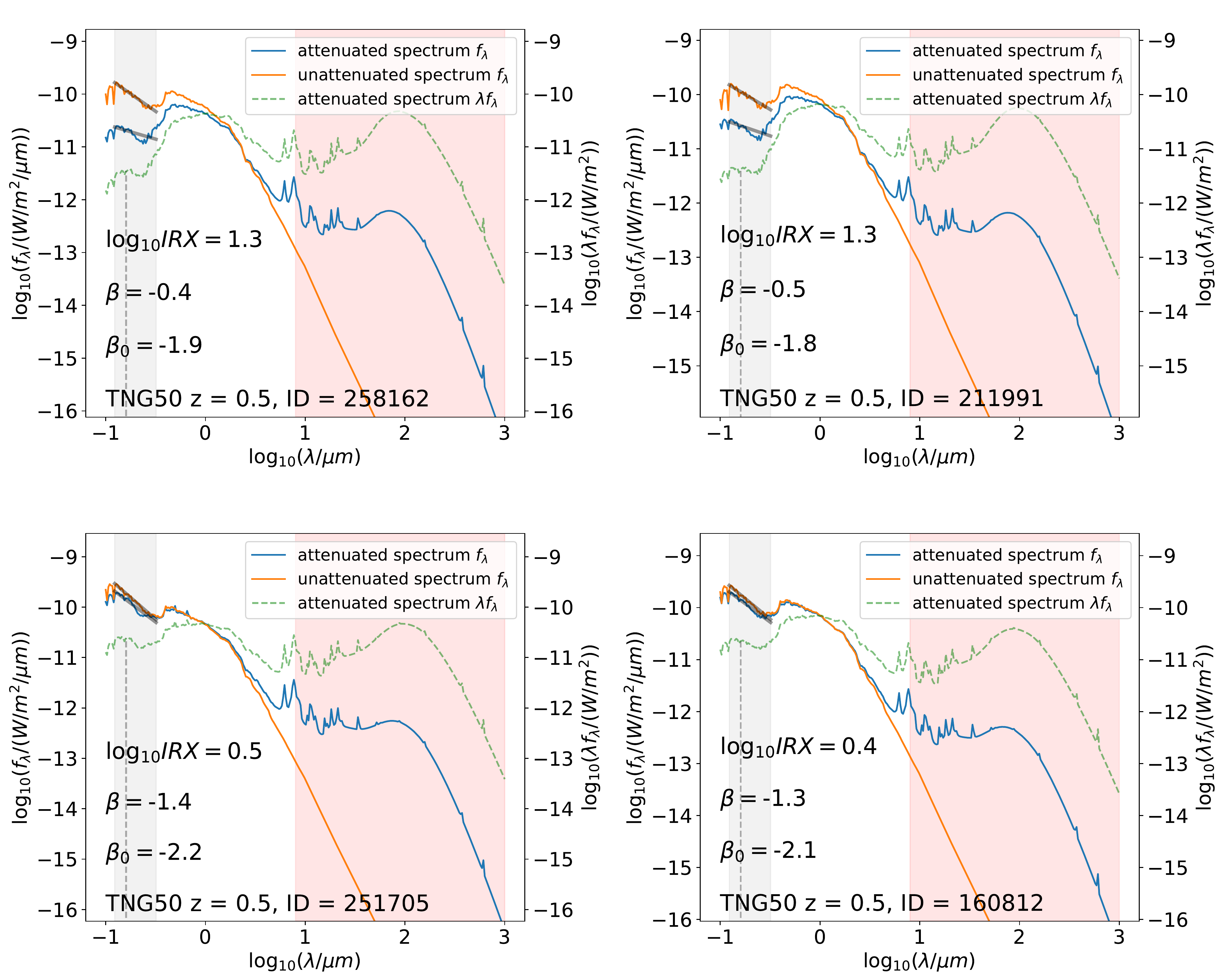}
    \caption{{\sc skirt} spectral flux density spectra of exemplary high mass galaxies of TNG50 at z = 0.5 (see their images in Figures \ref{fig:fig3} and \ref{fig:fig4}). For these spectra, we reran {\sc skirt} with a higher wavelength detail of 300 logarithmically distributed wavelength grid points and a higher photon package number of $10^7$ instead of $10^6$. The orange lines show the unattenuated spectrum, the blue lines show the attenuated spectrum, assuming Milky Way type dust properties. The grey shaded area marks the region in which we perform the fit to the attenuated (unattenuated) UV spectrum to attain its slope $\beta$ ($\beta_0$) (see section \ref{IRX-Beta Calculation}), which is indicated by the solid black transparent lines. The red shaded region shows the range over which we integrate to get $F_\text{IR}$. The green dashed line shows the flux density spectrum $F_\lambda = \lambda f_\lambda$ vs $\lambda$, from which we get $F_\text{UV}$ at $\lambda = \SI{0.16}{\micro\metre}$ (indicated by the black dashed line). Top row: Galaxies with relatively high IRX and high $\beta$. They are strongly attenuated, their UV slope changes significantly, and their UV spectral flux density is greatly decreased. Bottom row: Galaxies with relatively low IRX and low $\beta$. They are hardly attenuated at all, and show no significant change in UV slope, nor in UV spectral flux density. }
    \label{fig:fig2}
\end{figure*}

\subsubsection{Dust Modeling}\label{Dust Modelling}

{\sc skirt} offers the option of performing the radiative transfer calculations directly on a three--dimensional Voronoi mesh (\citealt{CampsVoro}), which makes it particularly well suited to the IllustrisTNG simulation suite, which is based on the {\sc arepo} code (\citealt{SpringelA}). We can directly reconstruct the gas distributions exactly as they are implemented in the hydrodynamic solver (except for cell gradients) in order to evolve the system. In practice, we define a cubical region around each inspected galaxy, with a side length of $15 r_{1/2}$ (with $r_{1/2}$ being the half stellar mass radius of the galaxy), to make sure that all of the parts of the dust distribution relevant to radiative transfer are captured. This ensures that even the contribution of very spatially extended dust is included in our simulations. The coordinates of the gas cells, which are in actuality mesh generating points, are used by {\sc skirt} to reconstruct the Voronoi mesh inside this volume using the {\sc voro++} open source library for computing Voronoi tesselations. Together with the density values for each gas cell, this fully describes a gas density distribution. 

We assume that the diffuse dust content of a galaxy is traced by the gas--phase metal distribution assigned to this galaxy. This means that we only include the gas cells which are gravitationally bound to the subhalo the galaxy is located in, all other gas cells' densities are set to zero beforehand. We add the condition that dust can only be present in a gas cell if the gas cell is either star forming ($\text{SFR} > 0$), or if its temperature is less than a typical threshold value ($T < T_\text{max}$, $T_\text{max} = 75000$ K). This means that we effectively exclude all gas cells that are too hot and have zero SFR as determined by TNG50. Cells that are star forming or that are below the temperature threshold are always considered to contain dust. In the TNG50 model, gas cells with a density above a threshold of $n_{H} \propto \SI{0.1}{\cm}^3 $ are considered star-forming and are stochastically converted into stellar particles (\citealt{Hernquist}, \citealt{PillepichTNG}).

We motivate these aforementioned choices by the fact that dust is rapidly destroyed in hot gas through thermal sputtering (\citealt{Guhata}).  Unfortunately, since TNG50 does not model dust directly, we cannot properly constrain $T_\text{max}$ using a physically motivated procedure. Other groups working with the {\sc eagle} simulations have chosen a value of $T_\text{max} = 8000$ K (\citealt{CampsSKIRT}, \citealt{Trayford}). We choose a much higher value so as to only eliminate the very hottest gas cells from potentially containing dust. We accomplish this by manually setting the metallicity of all gas cells that will contain no dust to zero, which in turn sets the dust density of these cells to zero.

Only a certain fraction of the metal content of a gas cell will be locked up in the form of dust. This fraction, the dust--to--metal ratio, is treated as a free parameter in this work. Orienting ourselves on other studies, we assume a constant fiducial value of this dust to metal ratio $f = 0.3$ (\citealt{CampsSKIRT}). Observational work (\citealt{Remy}, \citealt{DeVis}) and theoretical models (e.g., \citealt{McKinnon2017}, \citealt{Popping2}) suggest that this fraction evolves as a function of gas--phase metallicity, but the exact shape of this relation is still uncertain. It was also suggested that the dust--to--metal ratio is evolving with redshift, which is why \citealt{Vogelsberger2019} calibrated $f(z)$ depending on redshift, by comparing with observational results. However, for simplicity, we have assumed a fixed value for the factor $f$ in our fiducial model, which depends neither on the gas--phase metallicity nor on redshift. We have tested different values of this parameter to see the effect it has on the IRX--$\beta$ distribution of the galaxies (see section \ref{Varying Fraction}).

The dust density distribution is therefore derived from the TNG50 gas density distribution the following way:
\begin{equation}\label{eq:eq2}
    \rho_\text{dust} = \begin{cases}f\rho_\text{gas}Z & \text{if } (T_\text{gas} < 75000\text{ K})\ \rm{or}\ (\text{SFR}_\text{gas} > 0) \\
    0 & \text{if } (T_\text{gas} > 75000\text{ K})\ {\rm and}\ (\text{SFR}_\text{gas} = 0)
    \end{cases}
\end{equation}
where $\rho_\text{dust}$ is the dust density, $f$ is the dust--to--metal ratio, $\rho_\text{gas}$ is the gas density,  $Z$ is the gas metallicity, $T_\text{gas}$ is the gas temperature and $\text{SFR}_\text{gas}$ is the instantaneous SFR of the gas resolution elements. Note that the gas metallicity $Z$ is a factor in this equation - higher metallicity systems will hence have a higher dust--to--gas ratio via this definition.

{\sc skirt} offers various options to model the dust grain composition. As mentioned in Section~\ref{Introduction}, it is expected that the choice here will affect the resulting IRX--$\beta$ of our simulated galaxies. For our fiducial model, following \citealt{CampsSKIRT} and \citealt{Trayford}, we choose the \citealt{Zubko} multi--component dust mix, which models a composition of graphite, silicate and polycyclic aromatic hydrocarbon (PAH) grains, with various grain size bins for each grain type. The size distributions and the relative amount of the dust grains are chosen so as to recreate the properties of Milky Way type dust. We also later vary the dust type to recreate the dust of the Large Magellanic Cloud (LMC) and Small Magellanic Cloud (SMC), by employing a \citealt{Weingartner2018} grain size distribution for the LMC and SMC dust respectively. 

The dust of the molecular birth clouds mentioned in section \ref{Stellar Sources} is treated separately from the diffuse ISM dust, and is already included by the usage of the {\sc mappings-iii} models for star--forming regions before any radiative transfer calculations are being made. We assume that stellar particles older than 10 Myr will not be surrounded by birth clouds anymore, so their emission is only being absorbed and scattered by the ambient ISM (\citealt{Charlot_2000}), which is modelled using {\sc skirt}. 

We also repeat the {\sc skirt} run for each galaxy without the presence of any ISM dust and {\sc mappings-iii} birth clouds for the calculation of the intrinsic (unattenuated) UV--slopes $\beta_0$.  

\subsubsection{Virtual Instrument Setup}\label{Instrument}

Each galaxy is observed from an angle perpendicular to the xy--plane of the simulation box, from a distance of 10 Mpc. This leads to a random orientation of the galaxies within the sample with respect to our viewing angle. We choose a field of view equal to the box size of the dust distribution which is $15 r_{1/2}$, to again make sure that all of the parts relevant to radiative transfer in the galaxy are captured in the output spectra. This ensures that even very spatially extended UV emission, which might cause some IRX--$\beta$ deviations, is captured by our instrument. For the spatially resolved spectra, we choose a resolution of 100 by 100 pixels. These images will be used to investigate the spatial distribution of UV-- and IR--emission. This resolution is high enough to differentiate the galaxies' structures by eye.

\subsubsection{Performing the {\sc skirt} Runs}\label{Performing SKIRT runs}

{\sc skirt} is a Monte--Carlo radiative transfer code, meaning that the radiation field is represented as a discrete, large number of photon packages ("rays"). Typical numbers of rays per wavelength go from $10^6$ to $10^8$. In our fiducial model, we choose a value of $10^6$ rays per wavelength, which is a number that is capable of producing converged spectra. This means that increasing the number of photon packages would change the spectrum (i.e. the flux received at each wavelength) only on a sub--percent level, which is sufficient for the analysis performed in this study. Our fiducial wavelength grid is defined by a custom wavelength grid file, which contains a total of 59 wavelength grid points, with 30 evenly spaced grid points in the relevant UV--range $\SI{0.123}{\micro \metre} < \lambda < \SI{0.32}{\micro \metre}$, and 20 evenly spaced grid points in the IR-range $\SI{8}{\micro \metre} < \lambda < \SI{1000}{\micro \metre}$, and 9 grid points that are evenly spaced outside of these ranges, so that the whole grid spans the range $\SI{0.1}{\micro \metre} < \lambda < \SI{1001}{\micro \metre}$. This resolution is sufficient both for calculating the UV--slope $\beta$ and integrating the total IR--emission, even though very fine emission lines from the {\sc mappings-iii} models are not resolved. We performed a test run with a narrower spaced wavelength grid and found no significant deviations from the results produced by the fiducial model. 

\subsubsection{{\sc skirt} Output}

\begin{figure*}
    \centering
    \includegraphics[width = 0.999\hsize]{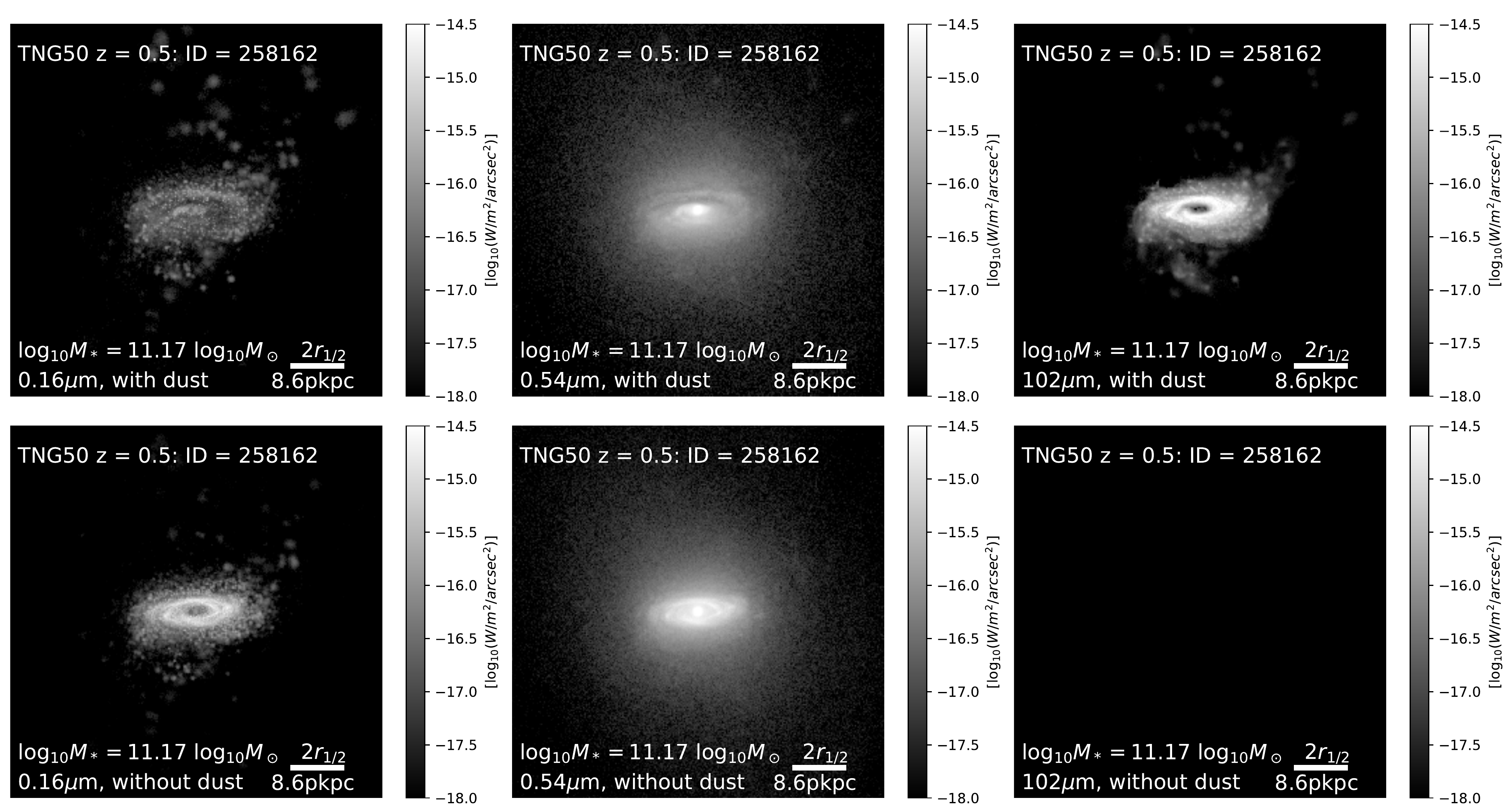}
    \includegraphics[width = 0.999\hsize]{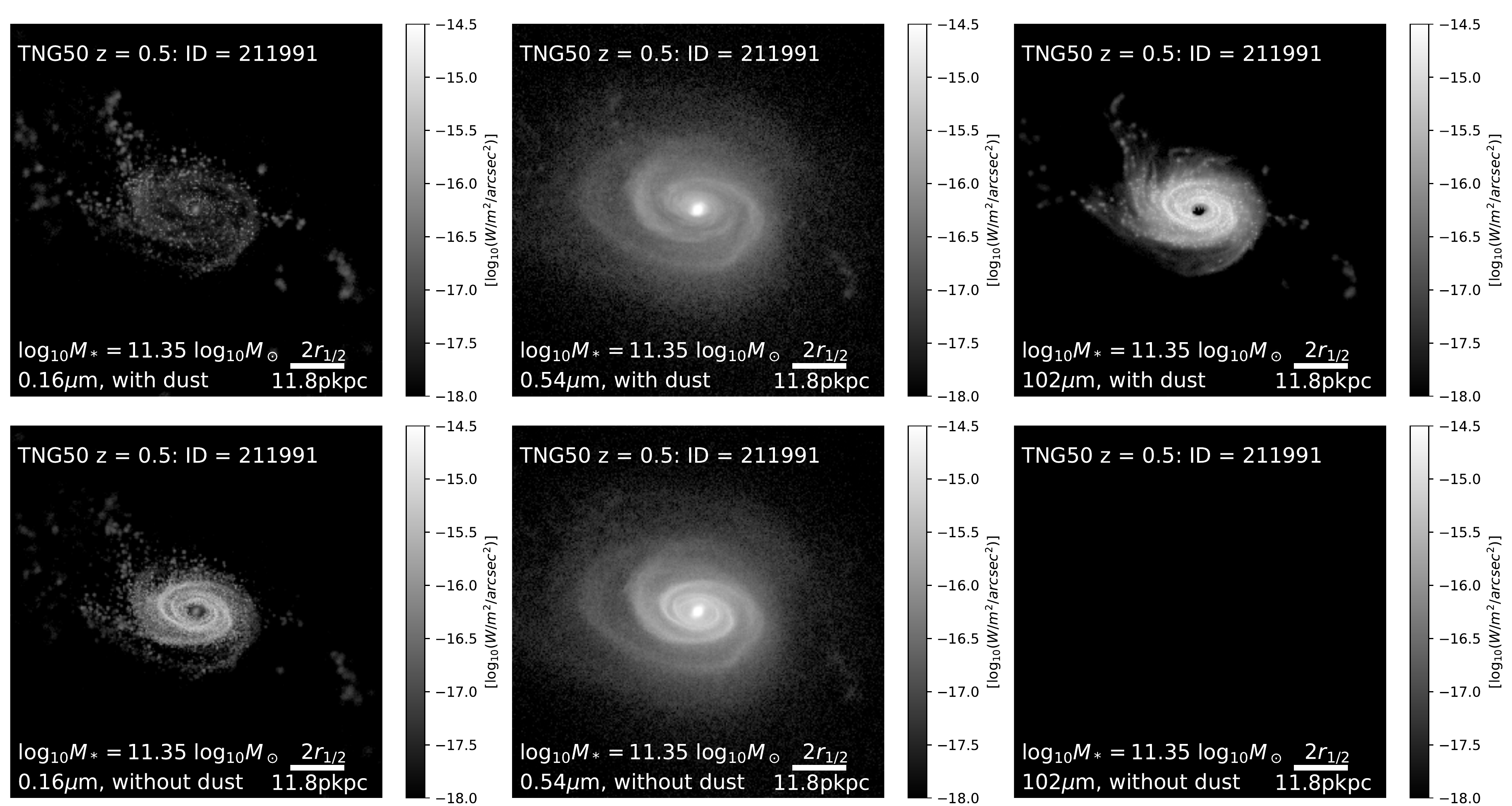}
    \caption{We show selected slices of the {\sc skirt} output data cubes for two high mass galaxies of TNG50 at $z = 0.5$ with high IRX and high $\beta$ ($\approx 10^{10-11} \MSUN$): see their spectra in Figure~\ref{fig:fig2}, top row). For these images we reran {\sc skirt} with 300 logarithmically distributed wavelength grid points, a higher photon package number of $10^7$ instead of the fiducial $10^6$ and a 500 $\times$ 500 pixel image resolution. For each galaxy, the respective top row shows the result when dust is included in the radiative transfer calculations, the respective bottom row shows the result when dust is excluded. From left to right, we show the spatial distribution of the UV flux ($\SI{0.16}{\micro\meter}$), the V--band flux ($\SI{0.43}{\micro\meter}$), and the IR flux ($\SI{102}{\micro\meter}$) as detected by a virtual instrument. The white bar in each image denotes a spatial scale of twice the stellar half mass radius $r_{1/2}$ of the depicted galaxy. For these strongly attenuated galaxies, the spatial distribution of their flux does not seem to change significantly when including dust attenuation. Noticeable is the strong change in UV and IR flux.}
    \label{fig:fig3}
\end{figure*}

\begin{figure*}
    \centering
    \includegraphics[width = 0.999\hsize]{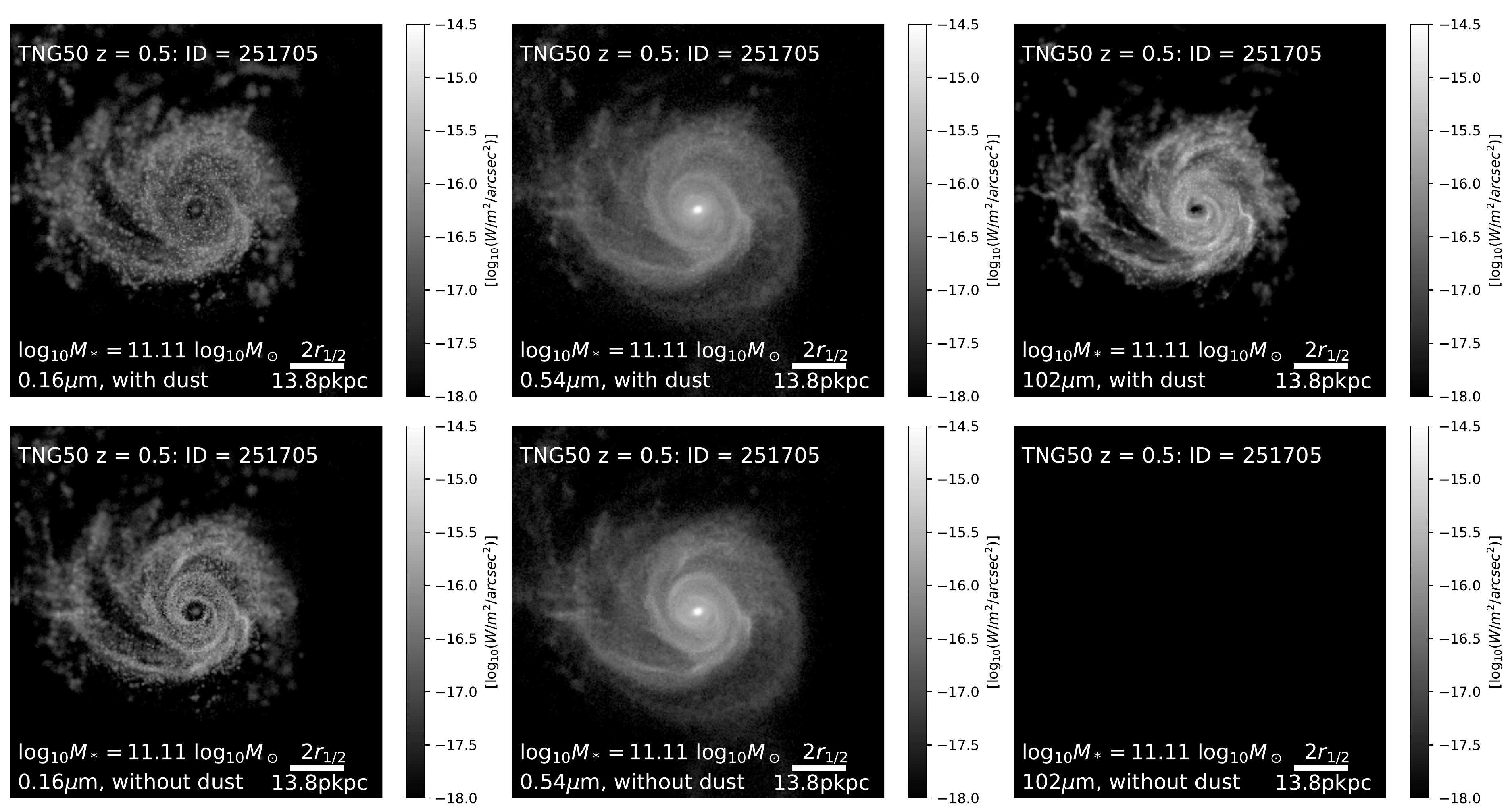}
    \includegraphics[width = 0.999\hsize]{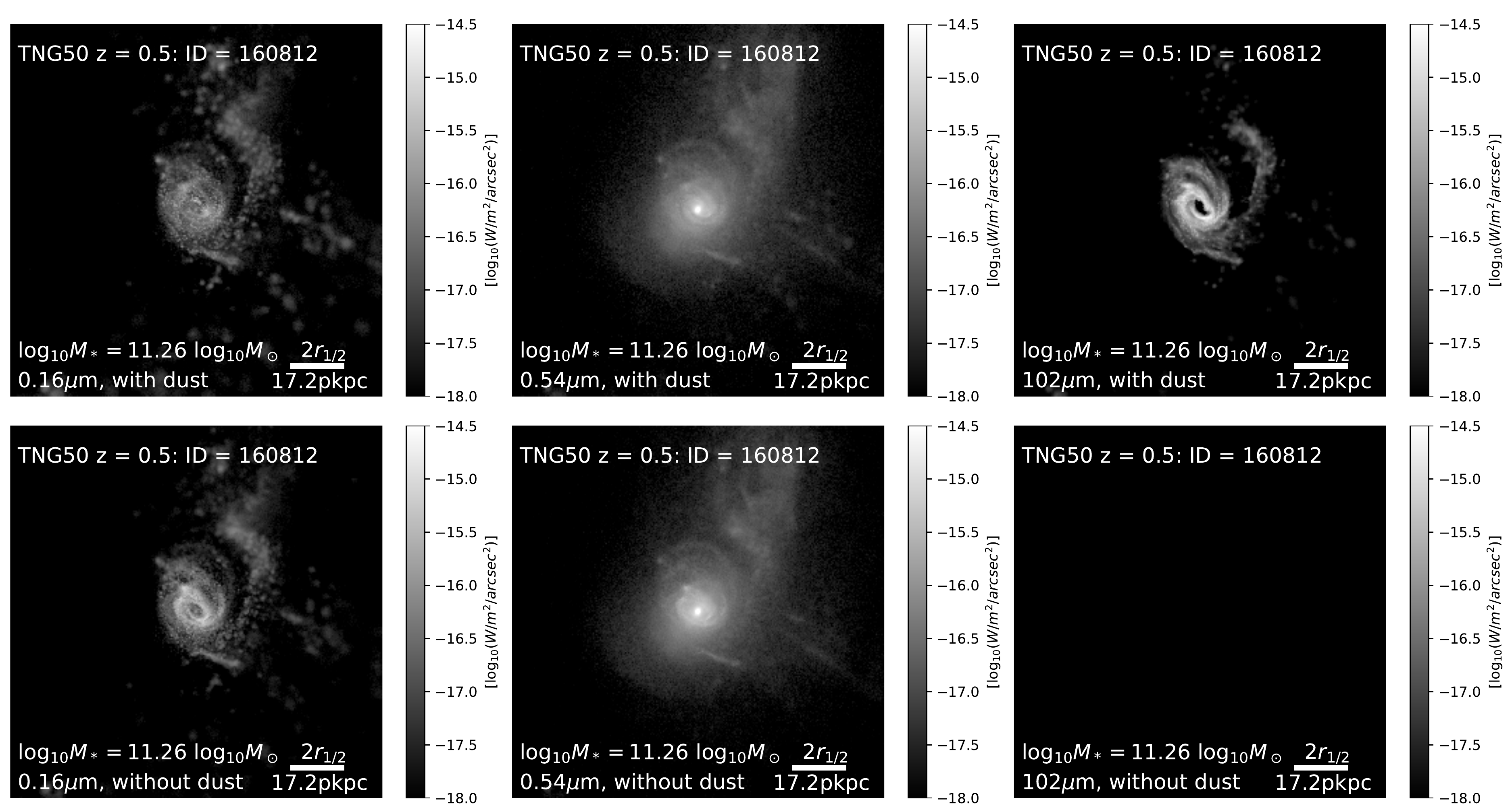}
    \caption{ We show selected slices of the {\sc skirt} output data cubes for two high mass galaxies of TNG50 at $z = 0.5$ with low IRX and low $\beta$ (see their spectra in Figure~\ref{fig:fig2}, bottom row).  Annotations are as in Figure~\ref{fig:fig3}. For these weakly attenuated galaxies, too, the spatial distribution of their flux does not change significantly when including dust attenuation. There is less decrease in UV flux when compared to the high IRX galaxies.}
    \label{fig:fig4}
\end{figure*}

\begin{figure*}
    \centering
    \includegraphics[width = 0.999\hsize]{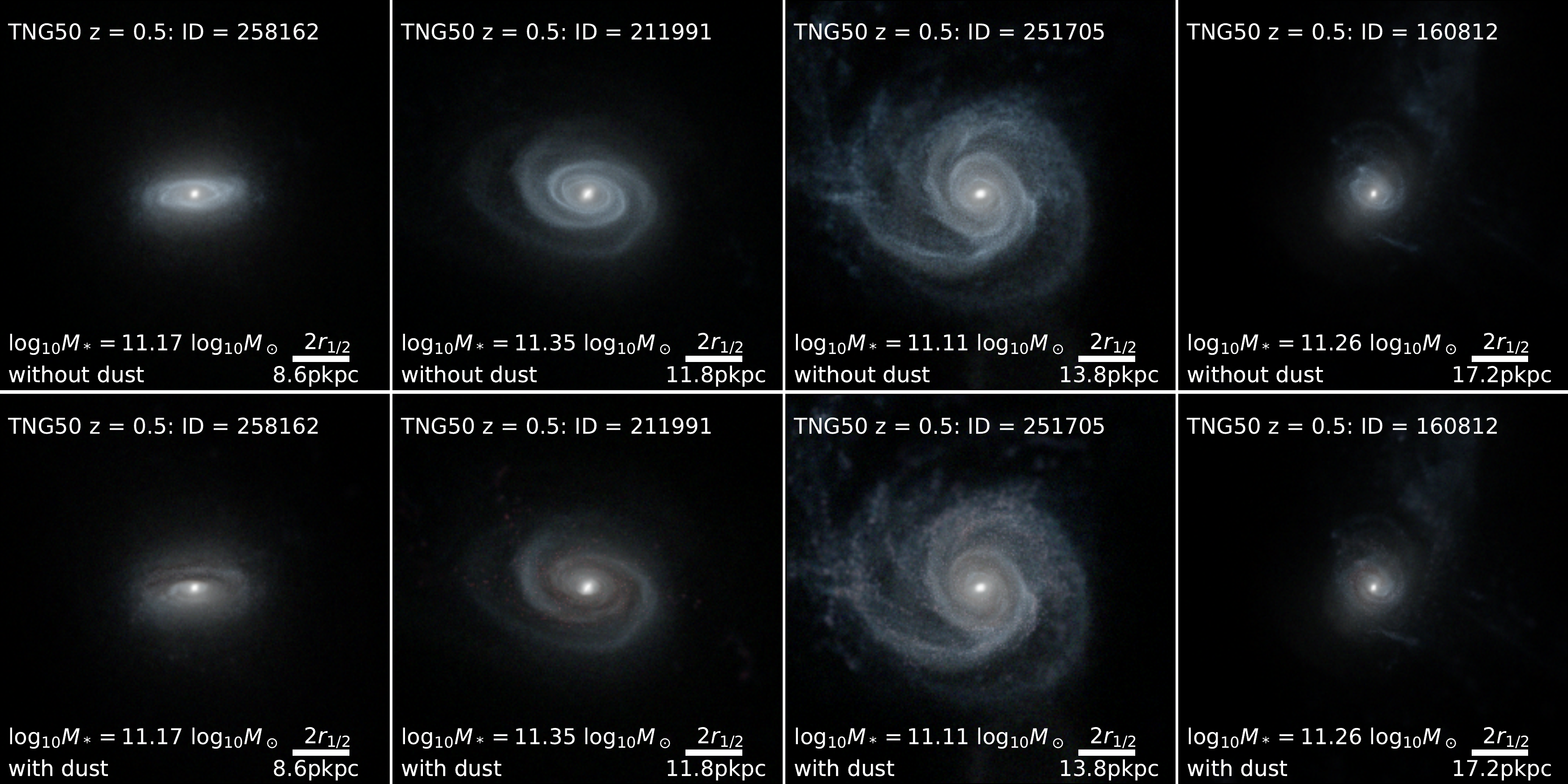}
    \caption{RGB images of the four high mass TNG50 galaxies of Figures~\ref{fig:fig3} and \ref{fig:fig4} at $z = 0.5$ produced from wavelength slices at $0.675\SI{}{\micro\metre}$ (red), $0.528 \SI{}{\micro\metre}$ (green) and $0.452 \SI{}{\micro\metre}$ (blue). The effect of dust on the perceived flux is clearly visible - many regions with high concentrations of young stars are covered by dust clouds when accounting for dust by application of radiative transfer models, affecting the resulting UV spectrum (see also Figure~\ref{fig:fig2}). Note also the reddening of some individual stellar particles due to the inclusion of {\sc MAPPINGS-III} spectra for very young stars, caused by effectively creating birth clouds in these star forming regions.}
    \label{fig:rgbimages}
\end{figure*}

The output of the {\sc skirt} runs are (i) full SEDs of each galaxy (given as flux density $F_\lambda = \lambda f_\lambda$ in $\SI{}{\watt/\metre^2}$ vs wavelength $\lambda$ in $\SI{}{\micro\metre}$) (see Figure~\ref{fig:fig2}), and (ii) data cubes containing resolved $100\times100$ pixel intensity maps for each wavelength of the specified wavelength grid, in units of $\SI{}{\watt/\metre^2/\text{arcsec}^2}$ (see Figures~\ref{fig:fig3} and \ref{fig:fig4}) which can then be combined, e.g. in order to create RGB images (see Figure~\ref{fig:rgbimages}). One can clearly see the effect that dust has on the appearance of galaxies at different wavelengths.

Summarizing, {\sc skirt} gives us the freedom to setup the physical models and the numerical specifications of the Monte--Carlo simulation. For the physical models, we have to specify a number of input parameters, which are either taken from the TNG50 output, from the literature, or they can be calibrated, or taken as a free parameter. For a summary of our {\sc skirt} setup, and a comparison to setups of other works coupling the IllustrisTNG model to {\sc skirt}, see Table \ref{tab:SKIRTset}. For an overview of the adopted {\sc skirt} input parameters and their origin please see Table \ref{tab:SKIRT}. 

\subsection{Measuring IRX and Beta}\label{IRX-Beta Calculation}

The infrared excess IRX is defined as the ratio between the IR luminosity and the UV luminosity and hence quantifies the amount of dust obscured emission from the galaxy:
  \begin{equation}
      \text{IRX} = \frac{L_\text{IR}}{L_\text{UV}},
  \end{equation}
where the IR luminosity $L_\text{IR}$ (the flux density $F_\text{IR}$) is attained from integrating the spectral luminosity $L_\lambda$ (the spectral flux density $f_\lambda$) over the wavelength range $\SI{8}{\micro\metre} < \lambda < \SI{1000}{\micro\metre}$, and the UV luminosity $L_\text{UV}$ (flux density $F_\text{UV}$) is the luminosity (flux density) at 1600 \AA. In this work we take the UV luminosity (flux density) at $\lambda=\SI{0.1601}{\micro\metre}$, the value in our wavelength grid that lies closest to that. Larger amounts of dust imply larger values of IRX if the EDAC is fixed, namely at fixed dust column densities, dust grain compositions and ISM dust geometries.

$\beta$ is defined as the rest frame UV spectral slope of a galaxy (where we assume that the spectrum follows a power law):
  \begin{equation}
      f_\lambda \propto \lambda^\beta
  \end{equation}
where $f_\lambda$ is the spectral flux density in the in the UV range. $\beta$ is fitted to the spectrum in the wavelength range $\SI{0.123}{\micro\metre} < \lambda < \SI{0.32}{\micro\metre}$, which corresponds to UV radiation from O and B stars and ensures a broad wavelength coverage around the $\SI{2175}{\angstrom}$ dust feature (if present, as for example in Milky Way dust).  Typically the spectrum of a galaxy is such that $\beta$ is negative: younger stellar populations produce a spectrum with a more negative (or bluer) $\beta$ than older stellar populations. Dust affects a galaxy's spectrum by reddening it and thus by making the UV slope less negative.

\section{Results}\label{Results}

In the following section, we investigate the IRX--$\beta$ relation of galaxies as predicted by the TNG50 simulation.  We compare our sample to a set of reference relations: (i) the original \citealt{Meurer} relation for local starbursts, (ii) recent calibrations by \citealt{Overzier_2010} for the same local starbursts, (iii) the \citealt{Casey_2014} relation which is a fit to low-redshift galaxies, and includes a higher dynamic SFR range and also aperture-corrected data of heterogeneous samples, and \citealt{Pettini} data for SMC-like dust extinction curves. We note that some of these works use different methods for determining the UV--slope $\beta$, e.g. different fitting ranges and a differing number of photometric data points in the fitting range. For example, \citealt{Meurer} use a collection of fitting windows as defined by \citealt{Calzetti94} in the wavelength range from $\SI{0.13}{\micro \metre}$ to $\SI{0.26}{\micro \metre}$. These windows were chosen in such a way to exclude absorption features like the $\SI{2175}{\angstrom}$ Milky Way bump from the fitting procedure. \citealt{Casey_2014} employ a similar method, also using Calzetti windows but extending the fitting range up to $\SI{0.32}{\micro\metre}$. These differences in the fitting methodology can lead to slight changes in the resulting $\beta$ (see for example \citealt{Popping}), e.g. due to the presence of the Milky Way bump feature. Nevertheless, when applying the method of \citealt{Calzetti94} to our data we find the overall relationship of galaxies across different redshifts to be similar to our fiducial model. More quantitatively, for a Milky Way dust composition, we find a median difference of $\Delta \beta \approx 0.1$ when comparing our method to the fitting method adopted by \citealt{Calzetti94}. The difference vanishes when adopting an SMC dust composition.

\begin{figure*}
\centering
  \includegraphics[width = 0.9\hsize]{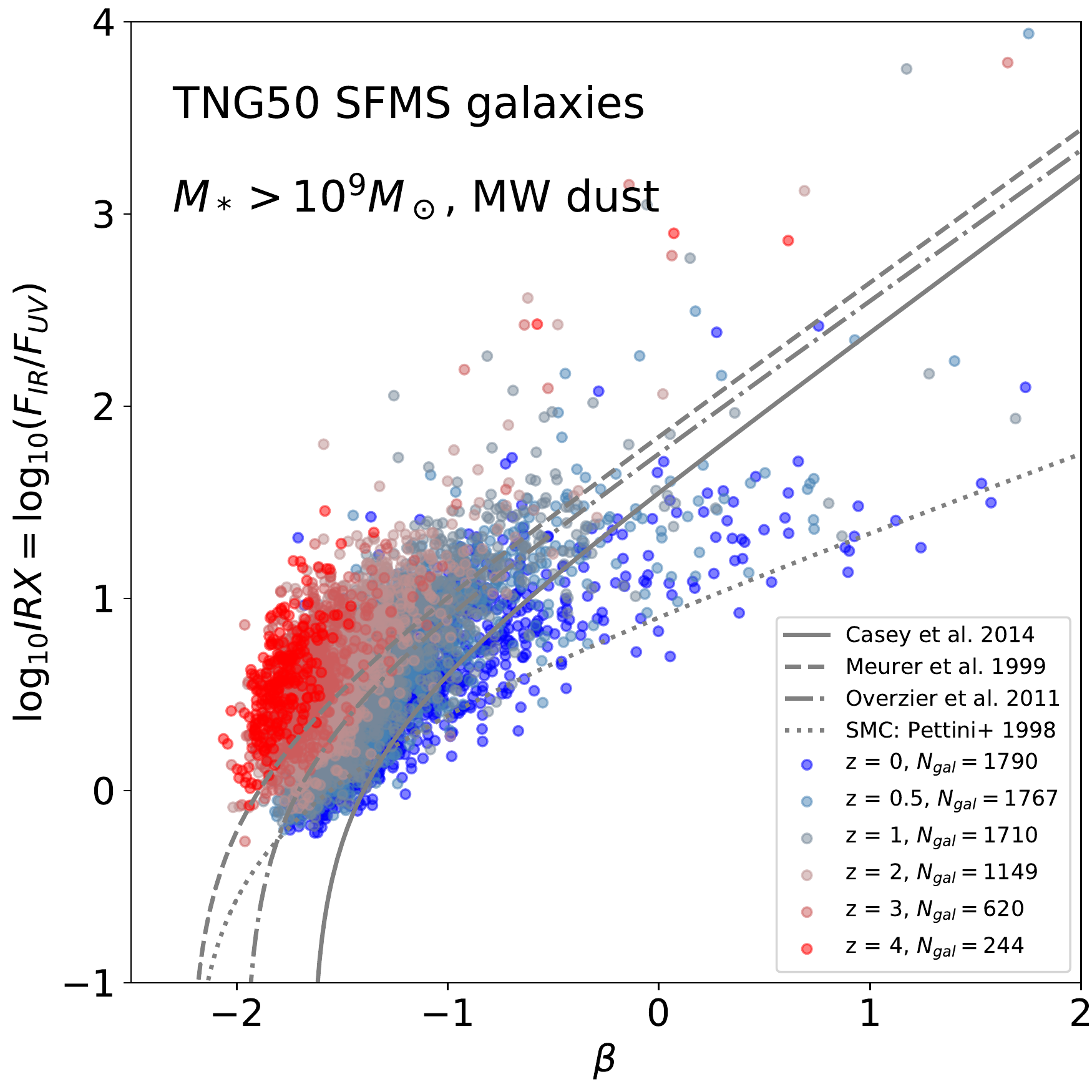}
  \caption{The IRX--$\beta$ distribution of the complete sample of selected TNG50 star-forming galaxies at all investigated redshifts. The lines show the various reference relations we compare our result to; colored circles represent galaxies from our selection in the $z =0-4$ redshift range, with $N_\text{gal}$ denoting the selected and depicted number of objects. The bulk of the galaxies agrees with the reference relations, the best agreement being with the Overzier curve (\protect\citealt{Overzier_2010}). We uncover a redshift trend: in general, at a given IRX, high--redshift galaxies have lower $\beta$ than low--redshift galaxies (see also Figure~\ref{fig:fig7},  top left panel). This indicates that high--redshift galaxies follow a different IRX--$\beta$ relation from low--redshift galaxies, under the assumption that the type of dust and dust-to-metal ratio do not change over cosmic time. TNG50 also predicts a non--negligible intrinsic scatter of the IRX--$\beta$ relation, for example at redshift 4, where most galaxies have a $\beta$ close to -2, their $\log_{10}\text{IRX}$ ranges from 0 to 1.}
\label{fig:fig6}
\end{figure*}

\begin{figure}
\centering
  \includegraphics[width = 0.9\hsize]{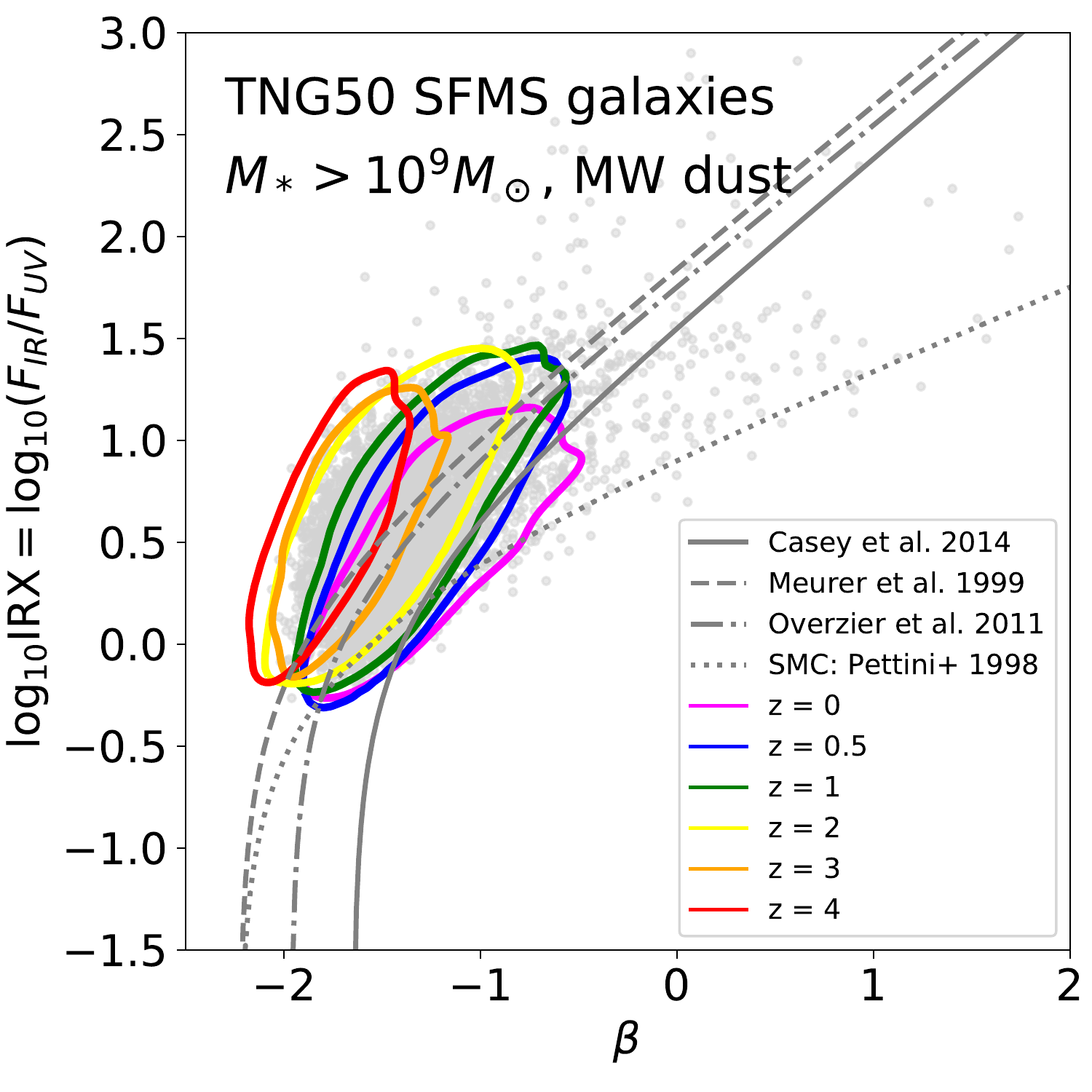}
  \includegraphics[width = 0.9\hsize]{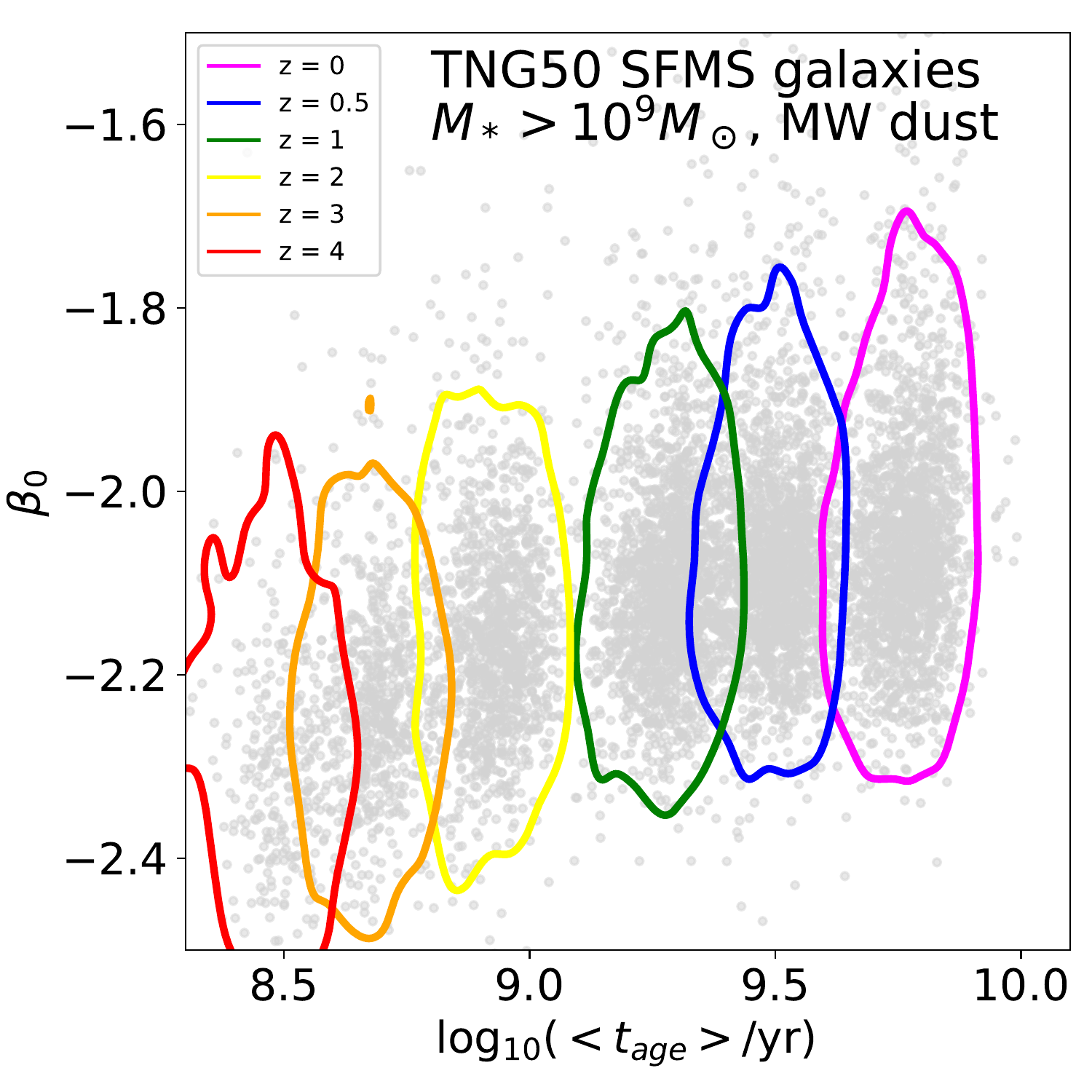}
  \includegraphics[width = 0.9\hsize]{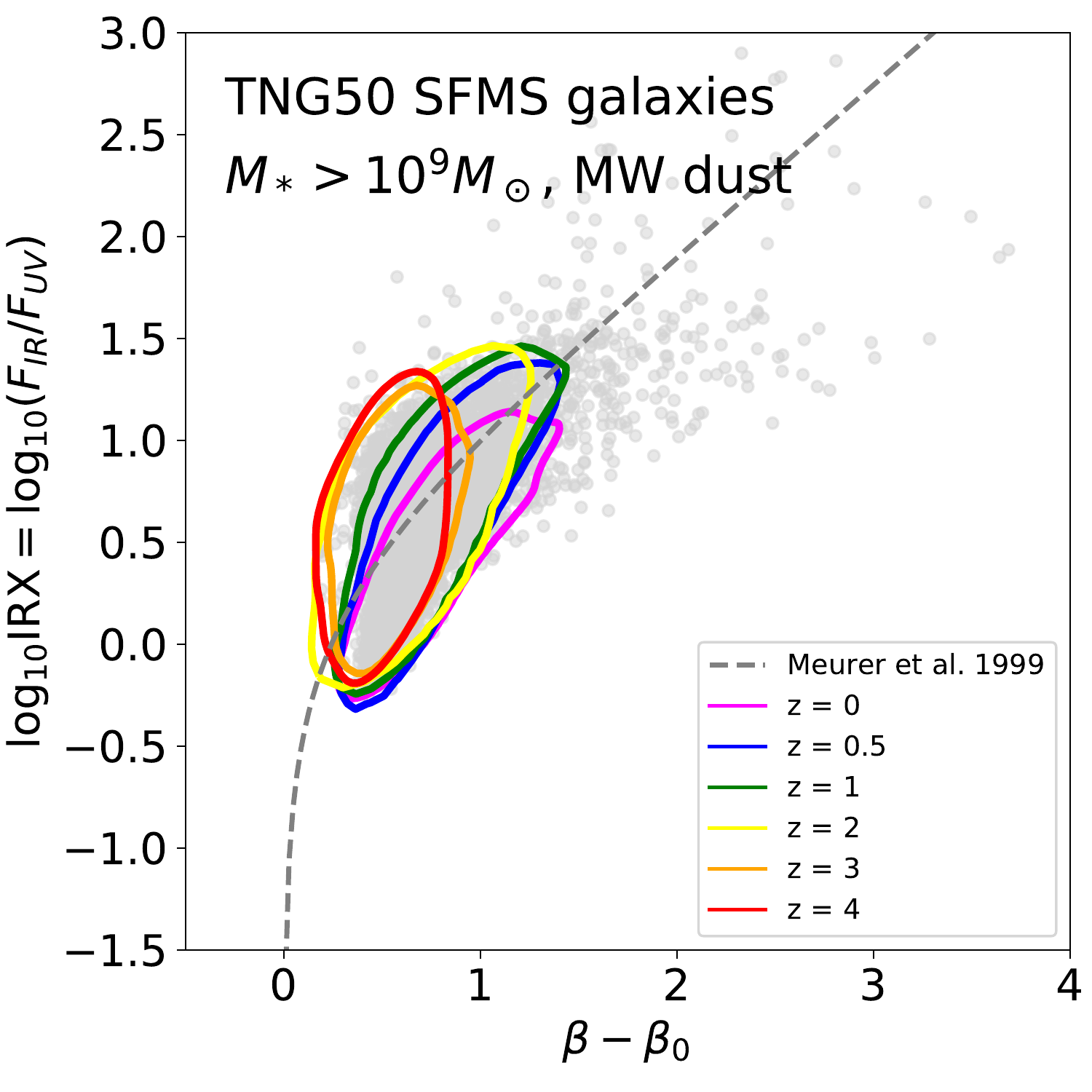}
  \caption{Top: redshift trend in IRX--$\beta$ from TNG50 galaxies. The distribution of the complete galaxy sample at all redshifts is shown by grey datapoints, the colored contours highlight the galaxy distributions and locations at each redshift (contours include 95\% of the population). At $z < 2$, galaxies generally follow the reference relations, while at $z > 2$, given IRX, galaxies are typically found at lower $\beta$ than what is predicted by the local reference relations. Caption continues in next column.}
\label{fig:fig7}
\end{figure}

\addtocounter{figure}{-1}
\begin{figure}
    \caption{Middle: galaxies' intrinsic UV--slopes $\beta_0$ against their mass weighted mean stellar population ages $\langle t_\text{age}\rangle$.
    The intrinsic slope $\beta_0$ correlates with the stellar population age: the younger the stellar population, the lower is $\beta_0$. High--redshift galaxies tend to have younger stellar populations than low--redshift galaxies and hence they have lower $\beta_0$. 
    Bottom: Effects of accounting for differences in stellar population ages and stellar metallicities across redshifts: we plot IRX against $\beta-\beta_0$, which is now a direct measure for the amount of UV--attenuation. This serves as a first order correction to redshift trends in IRX--$\beta$. The overall IRX--($\beta-\beta_0$) distribution is tighter, and the redshift dependent shift along $\beta_0$ has been reduced. 
    The high--redshift galaxies appear to still follow a steeper curve in the IRX--$\beta$ plane. We suggest that this is caused by systematically different stellar--to--dust geometries, manifesting themselves in different EDACs.}
\end{figure}

\subsection{IRX--Beta of TNG50 Galaxies}

Figure \ref{fig:fig6} shows a scatter plot of the IRX--$\beta$ plane for all galaxies in our sample, color coded by redshift. The bulk of the sample lies very close to the reference relations, and seems to be best approximated by the \citealt{Overzier_2010} relation (we demonstrate this in practice in the Discussion section). 

A typical galaxy in our sample has a $\beta$ between $-2$ and $0$, and a $\log_{10}\text{IRX}$ between 0 and $1.5$. We find that higher redshift galaxies lie towards lower $\beta$ at the same IRX: at $z = 4$, the median $\beta$ is $-1.80$, while the median $\beta$ at $z=0$ is $-1.36$ (see also Figure~\ref{fig:fig7}, top panel). The median $\log_{10}\text{IRX}$ is close to 0.5 for all redshifts. High--redshift galaxies seem to follow a steeper IRX--$\beta$ relation than low--redshift galaxies, with less scatter in $\beta$: the standard deviation of $\beta$ at $z=0$ is $\sigma_\beta (z=0) = 0.43$, while at $z = 4$ it is only $\sigma_\beta (z=4) = 0.31$. When compared to the scatter that has been found in observations (e.g. \citealt{Casey_2014}), the total scatter in our sample is smaller: at its worst, there is only a scatter of about 0.5 dex in IRX and a scatter of 1 in $\beta$ (compared to a scatter of ca. 1 dex in IRX and 1 in $\beta$ in \citealt{Casey_2014}). 

These deviations can have multiple causes: e.g. inclusion or not of observational measurement uncertainties, differences in the sample selections, variations in stellar population age and variations in the EDAC: see section \ref{On Scatter} for an elaboration on this. In the following, we will present results that help pinpoint those possible causes for the scatter and the redshift dependent trends in IRX-$\beta$ predicted by our model.

\subsection{Accounting for the Intrinsic UV--Slope}\label{Accounting}
Figure~\ref{fig:fig7}, top panel, shows explicitly that, according to TNG50 and assuming a Milky Way like dust and a fixed dust-to-metal ratio throughout, the IRX--$\beta$ steepens and gets shifted to lower $\beta$ at higher redshifts.

Differences in stellar population age can be a plausible source for scatter in the IRX--$\beta$ relation and for the deviations from a local shape. As the stellar population age increases, the intrinsic UV--slope ($\beta_0$) of the stellar population becomes redder (i.e. a less negative UV slope). This naturally creates scatter in $\beta$, without even invoking dust absorption.

For all the galaxies in our sample, their intrinsic UV--slopes $\beta_0$ can be obtained by running completely dust--free radiative transfer simulations (also excluding birth clouds) on them, and then measuring the slopes of the resulting unattenuated UV--spectra. In our model, these intrinsic UV--slopes depend solely on the galaxies' stellar metallicities and stellar population ages. 

In the middle panel of Figure~\ref{fig:fig7}, we plot the $\beta_0$ of all galaxies in our sample against their mass--weighted average stellar population ages $\langle t_{\rm age} \rangle$.  Higher redshift galaxies tend to have younger stellar populations, and this correlates with median intrinsic UV--spectra more "blue" (a more negative UV slope) when compared to low redshift UV--spectra. The median $\beta_0$ at $z=4$ is, for example $\beta_{0,~\rm{median}}(z=4) = -2.30$, while at $z = 0$, $\beta_{0,~\rm{median}}(z=0) = -2.07$. Apart from this systematic shift, we also see a significant scatter of $\beta_0$ across the galaxy population. At low redshifts, the spread in $\beta_0$ is also higher than at higher redshifts.  At $z = 4$, the standard deviation of $\beta_0$ is $\sigma_{\beta_0} (z = 4) = 0.10$ and at $z = 0$, it is $\sigma_{\beta_0} (z = 0) = 0.14$. 

In the bottom panel of Figure~\ref{fig:fig7} we now account for differences in stellar population age and stellar metallicities by plotting IRX against $\beta-\beta_0$. The colored contours show the distributions of the galaxies in this IRX--($\beta-\beta_0$) plane at each redshift. The value of $\beta-\beta_0$ is now a more direct measure of the amount of UV--attenuation in a galaxy, because it takes into account the natural fluctuations in the intrinsic UV--slopes $\beta_0$. The lower the value of $\beta-\beta_0$, the less "reddening" due to dust a galaxy has experienced. The redshift trend is now significantly reduced. Still, high--redshift galaxies have on average lower $\beta-\beta_0$ than low--redshift galaxies and, while the shift along $\beta$ has been reduced (weakly attenuated galaxies with low IRX values now coincide in the plot), the steeper slope at higher redshifts is even more apparent (for a fixed IRX we find a systematically different $\beta$) as a function of redshift. 

The IRX--$(\beta-\beta_0)$ distribution is tighter than the IRX--$\beta$ distribution. The remaining scatter in IRX at fixed redshift of about $\Delta \log_{10}\text{IRX} \approx 0.5$ and the systematically different slope for different redshifts in the IRX--$(\beta-\beta_0)$ relation can not be caused by differences in intrinsic UV--slopes (we corrected for $\beta_0$) and must therefore be caused by variations in the EDAC. A similar argument for a strong connection between the IRX--$\beta$ relation and the EDAC has already been made by \citealt{Salim_2019}. In our fiducial model this can only be related to variations in the stellar--to--dust geometry and the dust column density, as the dust grain composition is fixed globally to Milky Way type dust. In the following, we will investigate how different galaxy properties correlate with the location of galaxies in the IRX--$\beta$ plane and therefore with variations in the stellar--to--dust geometry.

\subsection{Examining Correlations in the IRX--Beta Plane}\label{Examining}
We investigate several physical quantities to see if they correlate with the galaxies' IRX--$\beta$ distributions. These include the star formation rates SFR, the average gas metallicity $\langle Z_\text{Gas} \rangle$, the specific star formation rate sSFR, the stellar mass $M_*$, the star formation efficiency $\text{SF}_\text{eff}$, and the gas mass $M_\text{Gas}$. All these quantities are measured within two times the stellar half mass radius of the modeled galaxies.
     
\begin{figure*}
    \centering
    \includegraphics[width = \hsize]{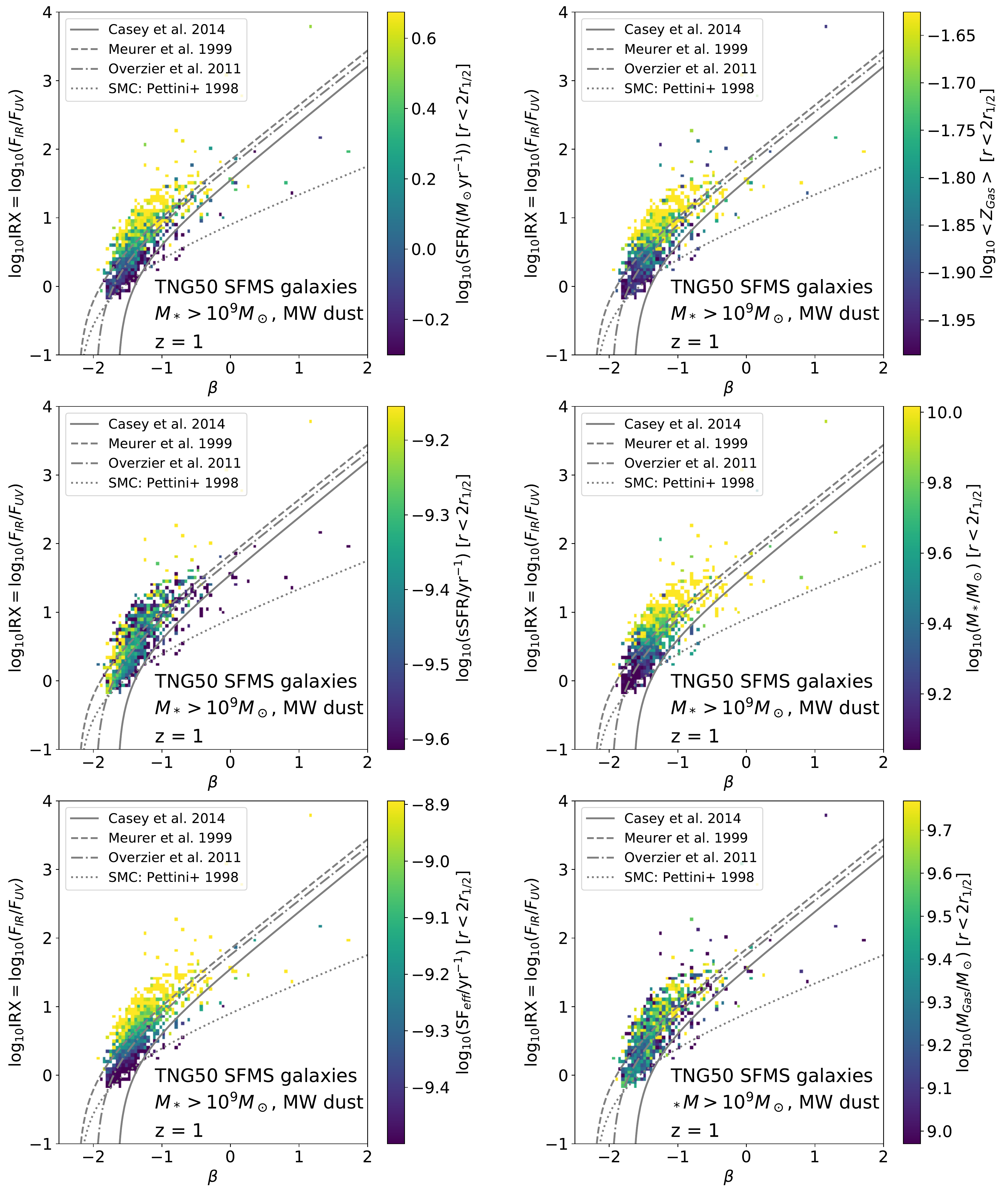}
    \caption{ Correlations of the IRX--$\beta$ distribution of the SFMS TNG50 galaxies at $z=1$ and  $M_* > 10^9 \MSUN$ with various physical galaxy properties, color coded. The IRX--$\beta$ plane is divided into 100 by 100 pixels, each containing a different number of galaxies (at least 1). In each pixel, we measure the median value of the physical quantity we investigate, and color code the pixel by this value. We study, from top to bottom, left to right: star formation rate; average gas metallicity; specific star formation rate; stellar mass; star formation efficiency; and gas mass, all within two times the stellar half mass radius.}
    \label{fig:fig9}
\end{figure*}

The panels in Figure~\ref{fig:fig9} show the IRX--$\beta$ distribution of TNG50 SFMS galaxies at $z = 1$ with $M_* > 10^9 \MSUN$. These plots are divided into 100 by 100 grid cells, the color of each grid cell corresponds to the median value of the physical property of interest of all the galaxies that lie inside of it in the IRX--$\beta$ plane. By plotting the IRX--$\beta$ distribution at fixed redshift, possible redshift trends that might be present in the whole sample across all redshifts are eliminated. We show $z=1$ as a representative redshift and we have checked that the findings below hold at all individual cosmic epochs between $z=0$ and $z=4$. Also, we found that most of the correlations in Fig. \ref{fig:fig9} (except from the correlation with sSFR) are qualitatively indistinguishable from those in the IRX--$(\beta-\beta_0)$ plane by visual comparison, suggesting that the main cause of the scatter in the IRX--$\beta$ plane is the variation of EDACs.

\begin{itemize}
    \item {\it Star Formation Rate (SFR)}. 
    Except for a handful of sources with $\beta > -1$ that fall below the locally derived reference relations, the IRX--$\beta$ relation depends on SFR (Figure~\ref{fig:fig9}, top left panel): at fixed IRX, higher SFR implies lower $\beta$ values. In fact, the layering due to different SFRs is somewhat parallel to the median  relation: in other words, at fixed $\beta$, higher SFR also implies larger IRX values. The overall effect would be accentuated if galaxy populations across cosmic epochs were considered, as the galaxy populations exhibit lower SFRs towards lower redshift at fixed mass.
 
    \item {\it Specific Star Formation Rate (sSFR)}. The left panel in the second row of Figure~\ref{fig:fig9} shows how the sSFR correlates with the IRX--$\beta$ distribution. When looking at  $z=1$, the correlation with sSFR is somewhat weaker, albeit still present, in comparison to e.g. the effects of SFR, and more prominent for galaxies with low IRX values:  higher sSFR galaxies tend to lie towards lower $\beta$ in the plane. This makes sense intuitively, as high sSFR galaxies contain on average larger amounts of younger stellar populations. Also in this case, the trends would be stronger if galaxies across many cosmic epochs were considered at once, as higher redshift galaxies have systematically higher sSFR and younger stellar populations.

\item {\it Stellar Mass}
We discern an almost one--to--one correlation between mass and IRX when showing the IRX--$\beta$ plane color coded by stellar mass (Figure~\ref{fig:fig9}, second row, right panel), with IRX increasing as a function of stellar mass. 
The trend at $z=1$ with stellar mass largely resembles the trend with SFR, which is a natural cause of looking at main--sequence galaxies at fixed redshift.

\item {\it Gas Mass and Star Formation Efficiency}
The star formation efficiency is defined as the star formation rate inside $2 r_{1/2}$, divided by the galaxy's gas mass inside $2 r_{1/2}$ (this includes a contribution from atomic, molecular, and ionized gas):
\begin{equation}
    \text{SF}_\text{eff} = \frac{\text{SFR}}{M_\text{Gas}}
\end{equation}
The higher this value, the more efficiently the gas of a galaxy is converted into stars, conversely a lower star formation efficiency corresponds to a galaxy that converts only little of its gas into stars. 

The correlation of the IRX--$\beta$ distribution with star formation efficiency is very strong (see Figure~\ref{fig:fig9}, third row, left). The color gradient appears to be perpendicular to the reference relations. In general, a higher star formation efficiency leads to a shift to higher IRX and lower $\beta$ with respect to the reference relations. 
We can attempt to dissect if this trend is driven by changes in gas mass or SFR by looking at these quantities independently. We already saw there is a clear gradient in the IRX--$\beta$ plane when color coding galaxies as a function of their SFR. There is a correlation present with gas mass, but it is less strong (see Figure~\ref{fig:fig9}, third row, right panel). The IRX of galaxies on average increases with their gas mass along the IRX--$\beta$ reference relation (i.e., they also have higher $\beta$). The gradient when color coding galaxies as a function of their $\text{SF}_\text{eff}$ resembles the gradient with SFR more than the gradient with gas mass, suggesting the SFR is the main parameter driving the gradient in $\text{SF}_\text{eff}$.

\item {\it Gas Metallicity}
The average gas metallicity of a galaxy is an indicator of how much dust each of its gas cells contain via Equation \ref{eq:eq2}. We find that IRX increases as a function of gas--phase metallicity, whereas $\beta$ correlates with gas--phase metallicity  less strongly (Figure~\ref{fig:fig9}, top right panel). As can be expected, an increased dust abundance results in a higher fraction of the emission being absorbed from the UV and re-emitted in the IR.

\end{itemize}

Note that the dependencies on star formation rate, gas metallicity, stellar mass, and star formation efficiency found in Figure~\ref{fig:fig9} remain similarly strong also in the IRX-($\beta-\beta_0$) plane (i.e. removing the effects of young stellar populations) while that with specific star formation rate is much weakened.




\section{Discussion}\label{Discussion}

In this work, we obtained the IRX--$\beta$ distribution of simulated galaxies of the TNG50 simulation, the highest resolution flagship run of the IllustrisTNG project. We performed radiative transfer calculations using {\sc skirt} to realistically model the effects of dust attenuation in these galaxies. We have focused on main--sequence galaxies at $z$ = 0, 0.5, 1, 2, 3 and 4, with stellar masses $M_* > 10^9 \MSUN$. 

\subsection{Interpretation of the Results}\label{Interpretation}

\subsubsection{The Intrinsic Scatter in IRX--Beta}\label{On Scatter}
The TNG50 galaxy population investigated in this study broadly agrees with the IRX--$\beta$ relations that have been observationally established for local galaxies. We see similar trends with SFR as \citealt{Casey_2014}, even though the latter report larger scatter than we find in TNG50, at low redshift. It should be noted from the onset that in this work we do not account for observational uncertainties that may be responsible for inflating the scatter in the observational data (e.g., when inferring the total infrared luminosities of galaxies). Nevertheless, we can speculate that a different amount of scatter can be due to sample selections: these may affect the observed data sets, but also the simulated one. For example, it could be that, because of the still limited cosmic volume covered by TNG50, we do not sample in TNG50 the most extremely star--forming galaxies (starburst, ultra luminous infrared galaxies, and dusty star--forming galaxies), which are rare. 
The lack of sources with high IRX in our simulations may also be the direct result of our choice for the dust--to--metal fraction. An increase in the dust--to--metal fraction from our fiducial value 0.3 will result in a higher $\beta$ and more so IRX (we find an increase by 0.5 dex when adopting the extreme scenario for local galaxies of a dust--to--metal fraction of 0.9, see Appendix \ref{app:fdtm}), naturally increasing the range in the predicted IRX of galaxies.

At $z < 1$ we find little scatter in the IRX--$\beta$ relation of simulated galaxies, similar to the scatter found in the observed IRX--$\beta$ relation for galaxies on the main--sequence. We suggest there are at least two reasons for the scatter. Firstly, part of the scatter is caused by variations in stellar population ages and stellar metallicities, leading to slightly different $\beta_0$, as we tentatively show in Figure~\ref{fig:fig9}, albeit for $z=1$. This introduces a scatter along $\beta$. Secondly, we speculate the remaining part of the galaxy-to-galaxy variation could be due to differences in the dust column density and dust geometry: these, together with dust grain composition (here kept invariant by choice across galaxies), determine different galaxy EDACs and can lead  
to further scatter in $\beta$ for a given IRX (see also \citealt{Salim_2019}). The predictions by our model at $z \geq2$ are in agreement with the results by \citealt{Bourne2017}, \citealt{Fudamoto2017}, \citealt{McLure}, and \citealt{Fudamoto2019} who find an IRX--$\beta$ relation shifted bluewards of the $z=0$ reference relations. Other works, on the other hand, \citep[e.g.,][]{Bouwens2016,Reddy, AM2016} find that the IRX--$\beta$ of normal star-forming galaxies is similar to the $z=0$ reference relations, or even further towards the red. 

The general agreement between our model predictions and observations is very encouraging. A more quantitative comparison will require additional steps to mimic the observational methodology, for instance by selecting galaxies in the exact same way as the aforementioned works. Additional scatter and uncertainty in the observations is introduced by assumptions about the dust temperature and other measurement errors (see e.g., the discussion on measuring $\beta$ in \citealt{Popping} and the discussion on dust temperature in \citealt{Narayanan}). In other cases the results are based on stacking \citep[e.g.,][]{Bouwens2016} rather than individual detections. These assumptions and approaches are not accounted for in our analysis. Furthermore, it should be kept in mind that we assigned Milky Way type dust to all simulated galaxies globally. In reality there might be a continuum of dust grain types varying throughout galaxies and over time. This would lead to a more spread out observed IRX--$\beta$ distribution with more scatter in observations. We will get back to this in Section \ref{Varying Composition}.

\subsubsection{Systematic Redshift Trends}\label{DisRedshift}

We discovered a systematic shift toward lower $\beta$ and steeper slopes in the IRX--$\beta$ plane with increasing redshift. We found that one driver behind the shift is the redshift evolution of the galaxies' intrinsic UV--slopes $\beta_0$, which evolves with redshift due to the following reason: galaxies at higher redshifts tend to have younger stellar populations. 
This causes their $\beta_0$ to be more negative, or more "blue" (see Figure~\ref{fig:fig7}) than those of low--redshift galaxies. 

After correcting for $\beta_0$, there are still non-negligible trends in redshift -- the slope of the IRX--$(\beta -\beta_0)$ relation is still steeper at high redshift than at lower redshfits. Again, we suggest this can be attributed to systematically different EDACs at higher redshifts, which can be driven by differences in the dust geometry or the dust column density (or both) across galaxies and at different cosmic epochs. The additional steepness of the high--redshift IRX--$\beta$ relation implies a weaker FUV--absorption in high--redshift galaxies due to geometry effects. One hypothetical explanation could be, for example, that the geometry of the dust distribution at high redshifts is less homogeneous, similar to a geometry of patches of dust or a dust screen with holes. This would cause the UV--slope $\beta$ to be dominated by unobscured UV--emission of young stars, while the IRX is dominated by dust emission of the dusty regions. We come back to this hypothesis in Section \ref{correl}.


We conclude that it is to be expected that the IRX--$\beta$ relation of Milky Way type SFMS galaxies with masses above $10^9 \MSUN$ is different at each redshift, with a trend toward lower $\beta$ and steeper slopes as we go to higher redshifts. This is driven by systematic differences in stellar population ages and star formation history, stellar metallicities, and possibly dust geometry. When inferring the SFR from such galaxies with the relations derived at $z \approx 0$, neglecting these systematic trends would lead to an underestimation of their IRX, which would lead to underestimations of star formation rates (see Section \ref{Quantifying}).

\subsubsection{Correlations between Galaxy Properties and IRX-Beta}\label{correl}

Since in the fiducial model the dust type has been fixed to Milky Way type dust for all galaxies, and we account for age related systematic shifts in $\beta$ by correcting for $\beta_0$, the remaining scatter  most certainly reflects the galaxy--to--galaxy property variations within the same population at given cosmic epoch, that in turn may determine differences in the geometry of the dust distribution around the light sources and differences in the dust abundance.
Indeed, most of the correlations found in this work are in place both for the entire sample of galaxies across redshifts, as well as at individual redshift snapshots.

We note that some of the investigated properties are not independent of each other. For the single redshift samples, the trend in SFR and in metallicity seems to be dominated by the trend in stellar mass (Figure~\ref{fig:fig9}, 2nd row, right) - higher mass galaxies on the main sequence will have higher SFRs and higher metallicities, and hence higher IRX. The sSFR accounts for mass trends in single redshift samples, because we divide by the stellar mass, and thus eliminate the inherent mass dependency of the SFR at fixed time. There, we do not see a correlation with IRX any more, but with $\beta$, which we suspect to be caused by different intrinsic UV--slopes $\beta_0$. The higher the sSFR, the lower is $\beta_0$ and hence $\beta$. We confirmed this suspicion by inspecting the corresponding IRX--($\beta-\beta_0$) plot, where this correlation vanishes.

There is a strong correlation between a galaxy's star formation efficiency and its position in the IRX--$\beta$ plane, perpendicular to the empirically derived reference relations. We saw that this is mostly driven by a correlation with SFR, rather than gas mass. This correlation further hints to the idea that the dust geometry of high--redshift galaxies might be less homogeneous compared to low--redshift galaxies. High star formation efficiencies can be achieved when the ISM gas is concentrated in star--forming regions. Since in our model the ISM dust is traced by the ISM gas, such a hypothetical scenario would lead to dust being concentrated in small patches of star--forming regions. Then, the UV--slope $\beta$ might be dominated by UV--emission from regions that contain only little dust, while the IRX might be dominated by emission from the dust that is centered on these highly star--forming regions, leading to quite high IRX values. This would give a plausible explanation as to why the FUV--attenuation is lower in high--redshift galaxies, leading to steeper curves in the IRX--$\beta$ plane. In the end, this has to be confirmed by a thorough investigation of the galaxies' EDACs themselves, and this could in principle be done with our simulation results. However, as this would surpass the scope of this study, we leave this open to future research.

The correlation between the average gas metallicity and the IRX--$\beta$ distribution can be explained as follows. Low metallicity galaxies contain less dust, and are therefore less opaque. This results in lower $\beta$ (less UV--emission is absorbed and scattered away) and lower IRX. 
We find that the IRX of galaxies increases as a function of stellar mass. At fixed redshift, $\beta$ also increases as a function of stellar mass. Higher mass SFMS galaxies generally have higher gas mass and dust--to--gas fractions (because they have higher gas metallicities), causing more dust obscuration and therefore higher IRX. 

Putting all of this together, we find that the location of different tracks in the IRX--$\beta$ plane (above and below the different reference relations) is best determined by the sSFR and the star--formation efficiency of galaxies, where the sSFR is a good indicator of variations in stellar population age and hence the shift along $\beta$, and the star--formation efficiency might be a good indicator of variations in the stellar--to--dust geometry and hence the change in slope. The infrared excess IRX on the other hand scales more strongly with stellar mass and gas-phase metallicity.

\begin{figure*}
    \centering
    \includegraphics[width = \hsize]{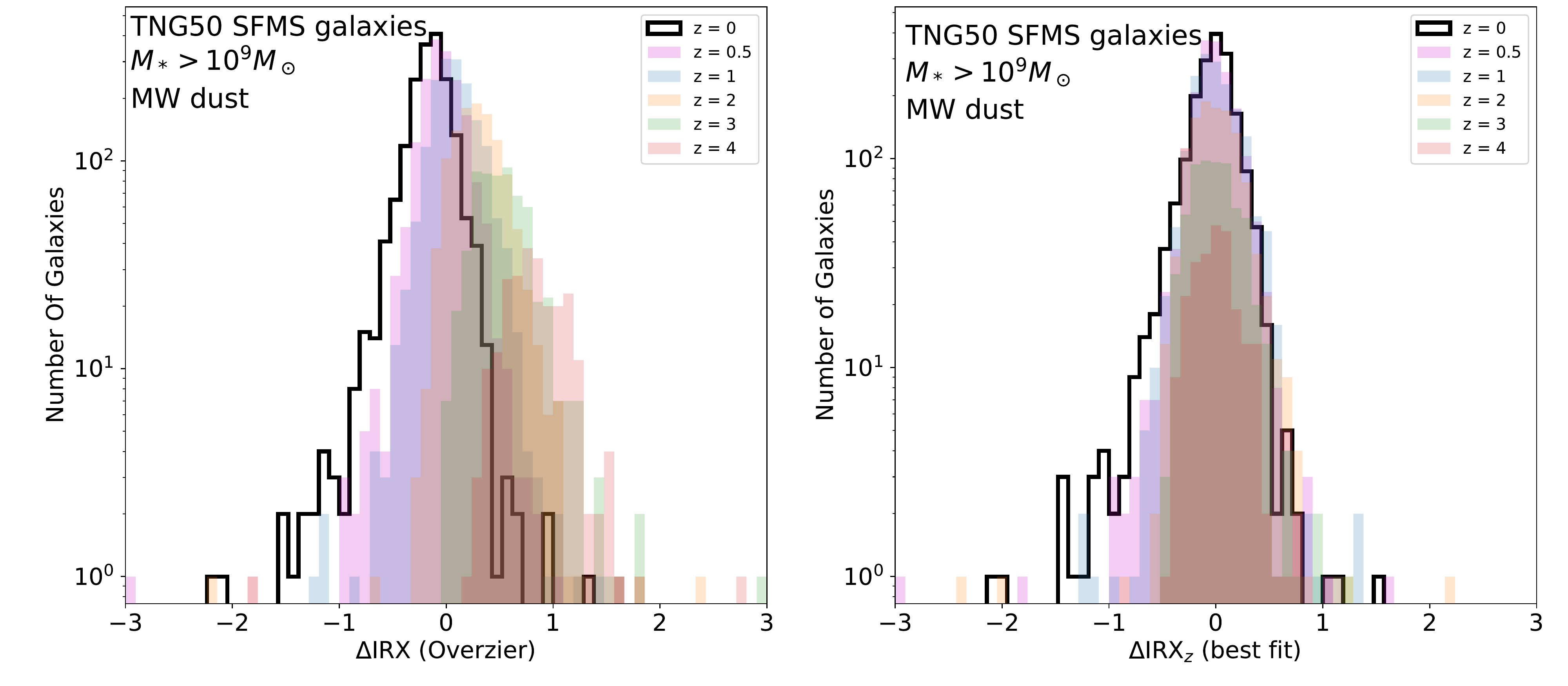}
    \caption{Left: A probability density histogram of $\Delta \text{IRX}$,  which is the difference between $\log_{10}\text{IRX}$ as obtained from our radiative transfer and TNG50 calculations and $\log_{10}\text{IRX}_\text{Overzier}$ as obtained from the Overzier's relation (\protect\citealt{Overzier_2010}), given $\beta$, binned as a function of redshift. The peak of the distribution moves to higher $\Delta \text{IRX}$ as we increase redshift, indicating that there is a redshift evolution in the IRX-$\beta$ relation. Right: We fit to our data at fixed redshift a modified version of the Overzier's relation, where we introduce a new parameter $\beta_z$ in Equation \ref{eq:eq3} in order to account for the systematic redshift trend. The panel shows the histogram of $\Delta \text{IRX}_z\text{(best fit)}$, which is now the difference between the predicted $\log_{10}\text{IRX}$ at fixed $\beta$, and the best--fit $\log_{10}\text{IRX}_{z}$. The redshift evolution is now accounted for in the new formula with all distributions peaking at 0.}
    \label{fig:fig13}
\end{figure*}

\begin{table}
    \centering
    \begin{tabular}{cccc}
    \hline
         z & $N_\text{total} $ & $\beta_z\text{(best fit)}$ & $N_{\Delta\text{IRX}_z\text{(best fit)} > 0.5}/N_\text{total}$\\
    \hline
         0 & 1790 & {-0.0941} & 0.01\\
         0.5 & 1767 & {-0.0165} & 0.01\\
         1 & 1710 & {0.0683} & 0.02\\
         2 & 1149 & {0.2272} & 0.03\\
         3 & 620 & {0.3472} & 0.02\\
         4 & 244 & {0.4696} & 0.03\\
     \hline
    \end{tabular}
    \caption{Results of the fit of the TNG50 data to a modified Overzier-like relation, where we introduce a new parameter $\beta_z$: see equation~(\ref{eq:eq3}). The best fit $\beta_z$ are listed in the third column, for TNG50 galaxies at a given redshift (first column) whose number is indicated in the second column. 
    The number of galaxies that strongly deviate from this modified relation ($N_{\Delta\text{IRX}_z\text{(best fit)} > 0.5}$, last column) is small, below 3 percent at any redshift. This implies that after accounting for a redshift shift, the locally observed relations hold for SFMS galaxies.}
    \label{tab:tab2}
\end{table}

\subsection{Observational Implications: a new redshift--dependent Fit for the IRX--$\beta$ Dust Attenuation Relation}
\label{Quantifying}
 The bulk of the galaxies modeled in this work have IRX and $\beta$ that agree well with the relation presented in \citealt{Overzier_2010}. We show this in Figure~\ref{fig:fig13}, where the mean offset between TNG50 galaxies across redshifts and the \citealt{Overzier_2010} relation for local galaxies read approximately 0.15 dex in IRX. This figure shows histograms of the differences between each galaxy's $\log_{10}\text{IRX}$ predicted by our model and the $\log_{10}\text{IRX}_\text{Overzier}$ obtained by applying the fit presented in \citealt{Overzier_2010} for the same $\beta$. Analog histograms for the fits presented by \citealt{Meurer} and \citealt{Casey_2014} (not shown) would give a worse agreement, with an average offset of 0.3 and 0.33 dex in IRX, respectively, at $z=0$.
 
Especially at $z \leq 0.5$ the modeled IRX as a function of $\beta$ is close to the IRX one would derive from the relation presented in \citealt{Overzier_2010}. The left hand side of the figure shows the histogram of $\Delta \text{IRX} = \log_{10}\text{IRX} - \log_{10}\text{IRX}_\text{Overzier}$ for discrete redshift bins. The peak of the distribution moves toward higher $\Delta \text{IRX}$ as we increase redshift. While most $z=0$ galaxies pile up around $\Delta \text{IRX} = -0.15$, the $z = 4$ galaxies have their peak at $\Delta \text{IRX} = 0.79$, implying that there is a redshift evolution in the IRX--$\beta$ relation (driven by variations in the intrinsic UV--slope and the EDAC as a function of redshift, as discussed in Section \ref{Accounting}). 
The fraction of galaxies that deviates from the Overzier et al. fit increases with increasing redshift, up to  50\% at $z= 3$ and even 90\% at $z=4$.

To account for the redshift trend, we fit a modified version of the \citealt{Overzier_2010} relation to our data at each redshift. With this fit, we aim to quantify the combined effects of systematic shift along $\beta$ and steepening of the slope of the IRX--$\beta$ relation due to varying $\beta_0$ and EDACs across redshifts. The combined effect of both of these contributions appears like a systematic shift toward lower $\beta$ with increasing redshift, to first order, with only a slight increase in slope in the IRX--$\beta$ relation. Thus, we limit ourselves to only one parameter in the modified fitting equation that incorporates both contributions from systematically varying $\beta_0$ and EDACs. To this end, we add a parameter $\beta_z$ as follows:
\begin{equation}\label{eq:eq3}
    \log_{10}\text{IRX}_z = \log_{10}1.68 + \log_{10}(10^{(0.4)(3.85 + 1.96(\beta + \beta_z))}-1)
\end{equation}
In this equation $\beta_z$ now allows for a shift of the \citealt{Overzier_2010} relation along $\beta$. The best fit $\beta_z$ at each redshift are listed in Table \ref{tab:tab2}. Note that at $z=0$ $\beta_z$ is not zero, which means we find a slightly different fit from Overzier et al. for local galaxies. We find that $\beta_z$ can then well be described as a linear function of redshift, such that:
\begin{equation}
    \beta_z(z) = 0.142z - 0.081
\end{equation}

In the right panel of Figure~\ref{fig:fig13}, we show the distribution of   $\Delta \text{IRX}_z\text{(best fit)} = \log_{10}\text{IRX} - \log_{10}\text{IRX}_{z}\text{(best fit)}$ at each redshift. The histograms are now centered on $\Delta \text{IRX}_z\text{(best fit)} = 0$.  After including $\beta_z$, the fraction of galaxies with  $\Delta \text{IRX}_z\text{(best fit)} > 0.5$ has dropped to $\sim 2-3$\% at all redshifts (see Table \ref{tab:tab2}).

The presented fit offers a direct approach for observers to improve the usage of the IRX--$\beta$ dust attenuation relation. We do note that the presented relation should be further tested observationally at redshifts up to $z = 4$, especially since in this work we assumed a uniform Milky Way type dust distribution. Beyond that, it is worthwhile to explore the validity of the presented relation at $z>4$.

\begin{figure*}
    \centering
    \includegraphics[width = \hsize]{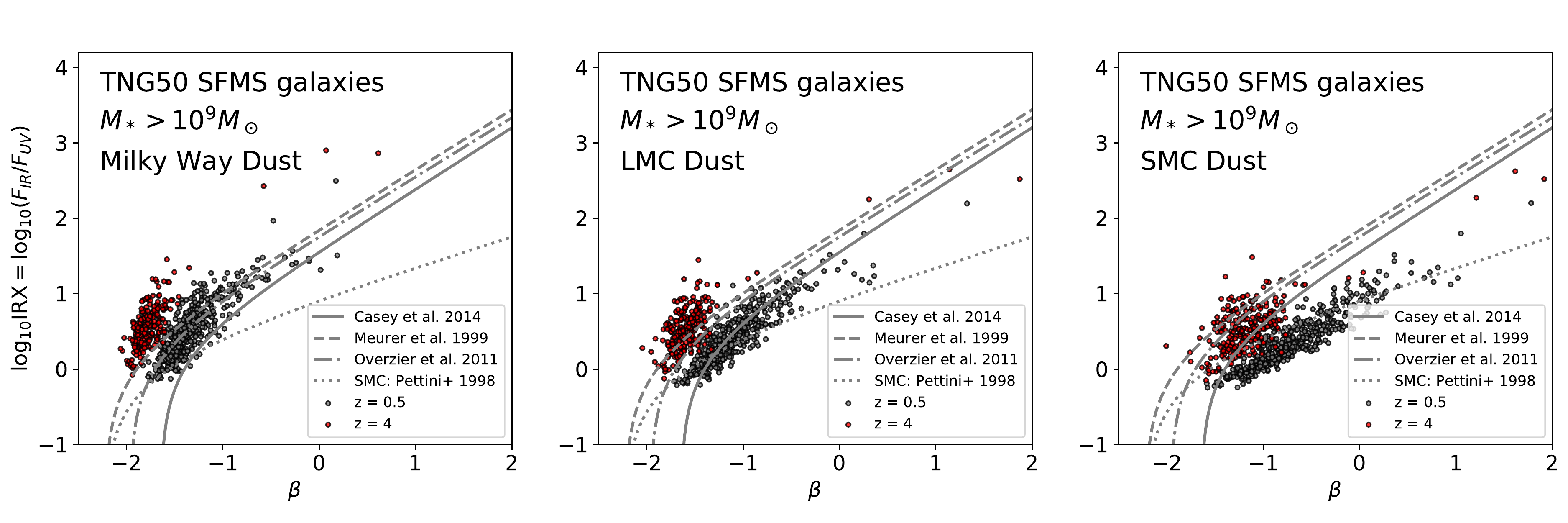}
    \caption{IRX-$\beta$ for a random subset of galaxies at $z=0.5$ and at $z=4$ for different assumptions of dust type. From left to right: Milky Way dust (as used throughout as fiducial choice) LMC, and SMC dust. As we vary the dust type to higher FUV absorbing dust, the slope in the IRX-$\beta$ relation flattens.}
    \label{fig:fig14}
\end{figure*}
\subsection{The Impact of Dust Composition}\label{Varying Composition}

The fiducial model of this work assigns \citealt{Zubko} Milky Way type galaxy dust to all galaxies. However, in reality it is plausible for dust types to be different across different galaxy types and across redshifts. It has been suggested, for example, that high-redshift galaxies  have dust more similar to SMC type dust, rather than Milky Way type dust \citep[e.g.,][]{Capak}. 
To further explore the effects of different dust types, we assign dust as it is present in the LMC and SMC to a subset of galaxies at $z = 0.5$ and $z=4$ and compare the IRX--$\beta$ distribution to the same subset of galaxies with Milky Way type dust (Figure \ref{fig:fig14}). 

LMC and SMC dust types universally decrease the slope of the IRX--$\beta$ relation, in comparison to the slope with Milky Way dust. The effect is stronger for a SMC dust composition. This is because of different amounts of far-UV extinction (FUV extinction) in the characteristic extinction curves for the dust types. LMC dust -- and even more so SMC dust -- have a higher FUV to NUV extinction ratio than Milky Way dust. This leads to a quicker change in UV--slope $\beta$ as the dust opacity is increased. Therefore, the galaxies do not follow the typical Milky Way type dust relations any longer, but have a "redder" $\beta$ for the same IRX. When employing SMC dust, the bulk of our subset of galaxies at $z=0.5$ follow the relation found by \citealt{Pettini} for SMC galaxies.

We also see a change in the IRX--$\beta$ relation for SMC dust galaxies at $z=4$ when varying the dust type. Unlike for $z=0.5$, the relation does not follow the \citealt{Pettini} trend, but is actually close to the $z=0$ reference relation for Milky Way type galaxies. In addition to the lower intrinsic UV--slopes at $z = 4$ causing a shift to lower $\beta$, we suggest the dust geometries at $z=4$ can balance out the "extra" reddening due to SMC dust (compared to Milky Way dust) in the EDAC, which moves the galaxies towards the $z=0$ reference relation for Milky Way type dust, rather than the \citealt{Pettini} relation for SMC dust. This demonstrates a degeneracy between the effects of stellar population age, dust geometry and different dust grain types on the location of galaxies in the IRX--$\beta$ plane.

Some observations have found that $z=2-3$ galaxies generally agree with the low redshift reference relations (or lie towards slightly redder $\beta$) and found no trend with redshift contrary to what is suggested by our models (\citealt{Reddy, Bouwens2016}). If the dust composition of these galaxies is indeed different from Milky Way type dust, the change in $\beta$ due to dust composition could balance the change in $\beta$ due to a young stellar population age and dust geometry in such a way that the galaxies fall close to the $z=0$ reference relation. If this scenario is true, it means that the use of the IRX--$\beta$ plane to study the dust type in different galaxies is hampered by the age/dust geometry/dust type degeneracy. This degeneracy can be broken by measuring the stellar population age to account for the intrinsic UV--slope, by measuring the geometric distribution of the ISM dust in relation to the stellar distribution and by measuring the dust grain composition of these galaxies.

\subsection{Comparison with Previous Work}
The IRX -- $\beta$ dust attenuation relation has been studied using hydrodynamical models in the literature before. In this subsection we discuss these efforts and contrast them against the results presented in this work.

\citealt{Safarzadeh} analyzed a set of 51 idealized hydrodynamical simulations of disk galaxies and mergers at $z=0$ and $z=2-3$, on which radiative transfer was performed in post-processing. They found that at $z=0$, galaxies are usually found close to the relation of \citealt{Meurer}, while at $z=2-3$, they lie above it even though they are not necessarily starbursts. They explain these deviations by disassociated UV-- and IR--emission in these galaxies. The dust type significantly impacts the slope of the IRX--$\beta$ relation. Lastly, they find that ULIRGs will deviate strongly toward high IRX. 

\citealt{Narayanan} performed zoom--in simulations of a number of galaxies down to $z=2$ (and $z=0$ in one case) and concluded that Milky Way like galaxies, with relatively young stellar populations and cospatial IR-- and UV--emitting regions lie near the standard relations such as the \citealt{Meurer} relation. They also find causes for substantial deviations: old stellar populations fall below the canonical relations, complex dust geometries lead to a deviation above the relations, and are common in high--redshift heavily star forming galaxies, SMC dust decreases the slope in IRX--$\beta$ and ULIRGs generally fall significantly above the reference relations. 

\citealt{Ma} analyzed 34 zoom--in simulations of $z>5$ galaxies.
They assumed SMC dust in their radiative transfer calculations, and confirmed a tight relation in agreement with the \citealt{Pettini} relation, despite the patchy dust geometry they find in their modeled galaxies. They also report that higher redshift galaxies move toward bluer $\beta$ at fixed UV due to younger stellar populations and also find that dust type affects the slope in the IRX--$\beta$ distribution as well. 

Our results are consistent with those obtained by previous theoretical efforts in the literature. However, this is the first time the IRX--$\beta$ relation is investigated through theoretical models for a homogeneous and large unbiased sample of many thousands of galaxies that are fully representative of SFMS galaxies from $z=0$ to $z=4$ in a full cosmological context, albeit volume and mass limited. This strengthens the applicability of our results, particularly for the redshift evolution in the IRX--$\beta$ relation due to stellar population age and dust geometry.

\subsection{Limitations}\label{Limitations}

Our galaxy sample is mass--limited: we only investigate galaxies with masses $M_* > 10^9 \MSUN$. We confirmed that the stellar masses of galaxies and their dust contents correlate with their IRX -- this in turn means that our sample is limited to a minimal value of $\beta-\beta_0$, which can be seen in the bottom panel of Fig. \ref{fig:fig7}, making it hard to discern the true behaviour of the high--redshift population with very little attenuation $\beta-\beta_0 \approx 0$.

Throughout the analysis, we have fixed the ISM dust type of each galaxy globally, which means that every gas cell of each whole galaxy contains the same type of dust (Milky Way dust). In reality, it is plausible to expect that galaxies consist of a mix of dust types in different regions, which change the resulting IRX--$\beta$ compared to our results. It is also probable that the dust content does not scale linearly with the gas metallicity, which is assumed in our model. 

The dust--to--gas fraction may vary across different galaxy regions and over time. For example,\citealt{Vogelsberger2019} obtain a strongly declining dust--to--metal ratio as a function of redshift by comparing the TNG dust-attenuated rest-frame UV luminosity functions to existing observational data: their dust--to--metal varies between almost 0.9 at $z=0$ to 0.1 at $z=6$. It is plausible that a varying dust--to--metal ratio would return a different redshift evolution of the IRX--$\beta$ plane than the one put forward here (see Appendix~\ref{Varying Fraction}). As the galaxy population at $z=4$ appears to follow a steeper IRX--$\beta$ relation, its offset from the $z=0$ population might be influenced by our choice of a fixed dust--to--metal fraction of $f = 0.3$. If at $z=4$ this fraction is indeed lower (e.g. $f \approx 0.1$) in reality, this would place the galaxies more close to the local reference relation, after accounting for $\beta_0$, reducing the discrepancy that may be caused by geometric effects on the EDAC.
In recent years, multiple groups have implemented the tracking of dust and the extinction curve of dust in hydrodynamical simulations \citep{McKinnon2016,McKinnon2017, McKinnon2018,Hou,Li2019,Vogelsberger2019dust}. These approaches are a promising avenue to avoid having to make assumptions about the type and amount of dust in modeled galaxies.

The effects of birth clouds on the attenuation of stellar emission has been modeled by employing {\sc mappings-iii} spectra for young stellar populations. This crude implementation of unresolved birth cloud attenuation limits the predictive power of our model, especially for high--redshift galaxies and starbursts that contain high fractions of young stars. A better treatment of birth clouds will require higher resolution simulations that resolve the ISM structure in these objects.

 Finally, we caution the reader not to over interpret some cases of TNG50 objects that appear as strong outliers from the average IRX--$\beta$ relation. As documented in Section 5.2 of \citealt{NelsonTNG}, not all entries in the {\sc subfind} catalogs should be considered `galaxies', as they may result from the fragmentation or collapse of gas within already formed galaxies. The criterion adopted in this work (non-vanishing stellar mass and a total dark matter mass fraction of at least 20 per cent, see Section~\ref{Galaxy Selection}) might not be sufficient to exclude all these objects from the analysis and it is expected for them to appear as outliers in most relations among galaxy physical properties.

\section{Summary and Outlook}\label{Summary}
We have performed radiative transfer calculations with the publicly-available software {\sc skirt} on star--forming galaxies at $0\leq z \leq 4$ taken from the TNG50 simulation. Our fiducial model adopts a universal Milky Way type dust and a dust--to--metal fraction of 0.3 throughout. It considers a volume--limited sample of galaxies above $10^9~\MSUN$ in stellar mass (more than ten thousand stellar particles) at six different redshifts (from $z = 0$ to $z=4$), corresponding to a total of 7280 galaxies. We have therefore quantified the evolution of the IRX--$\beta$ dust attenuation relation of galaxies and how it correlates with galaxy properties. We summarize the insight we gained from this work as follows:

\begin{itemize}
    
\item The bulk of the TNG50 SFMS galaxies at $z<1$ follows the reference relations for $z \approx 0$ by \citealt{Meurer}, \citealt{Overzier_2010} and \citealt{Casey_2014}. Only a small fraction ($\lesssim5$ per cent) of the low-redshift TNG50 SFMS galaxies  with stellar mass $M_* > 10^9 \MSUN$ deviate from the local relation found by \citealt{Overzier_2010} (with $\Delta\text{IRX} > 0.5$). This justifies the use of these relations at these redshifts to account for dust obscuration when IR data is missing (see Figure~\ref{fig:fig13} and Table \ref{tab:tab2}.)\\

\item More generally, our model predicts a systematic redshift dependent shift along $\beta$ in the IRX--$\beta$ plane, where higher redshift TNG50 SFMS galaxies tend to be shifted toward lower $\beta$. We find this is in part caused by a lower median intrinsic UV--slope $\beta_0$ due to younger stellar population ages and we speculate the remainder of the variation to be due to variations in the star--to--dust geometry, in turn manifesting themselves in different EDACs across redshifts -- a similar connection has already been made by \citealt{Salim_2019}.
These effects result in TNG50 galaxies at high redshifts being poorly described by an \citealt{Overzier_2010} relation. 
We therefore derived a new redshift--dependent version of the relation of \citealt{Overzier_2010}, such that: 
\begin{equation}
    \log_{10}\text{IRX}_z = \log_{10}1.68 + \log_{10}(10^{0.4(3.85 + 1.96(\beta + \beta_z(z)))}-1),
\end{equation}
where $\beta_z(z) = 0.142 z - 0.081$ in our model. This fit describes the IRX of galaxies at $0 \leq z \leq 4$ well, with only a few per cent of strong outliers (see Figs.~ \ref{fig:fig7}, \ref{fig:fig13} and Table \ref{tab:tab2}).\\

\item  Several physical properties determine the location of TNG50 galaxies on the IRX--$\beta$ plane. 
We find that IRX increases with larger stellar mass, gas--phase metallicity, SFR and star formation efficiency. The IRX--$\beta$ as a whole shifts towards bluer (i.e. more negative) $\beta$ for increasing sSFR.\\

\item Dust composition plays an important role for both the shape and evolution of the TNG50 IRX--$\beta$ relation. The IRX--$\beta$ relation for LMC and SMC type dust (with a higher FUV--to--NUV extinction ratio than MW type dust) is flatter than it is for Milky Way type dust. In certain cases (we demonstrate this for $z=4$), the effects of young stellar age (resulting in a blue intrinsic UV--slope) and possibly the effects of dust geometry variations (which can result in an effectively lower FUV attenuation) balance out the effects of SMC type dust (which increases $\beta$ to redder values compared to a typical Milky Way type of dust), such that a galaxy may appear to follow the reference local IRX--$\beta$ relation for galaxies with Milky Way type dust, despite having SMC type dust. Without additional knowledge on the stellar population properties and dust geometry of galaxies, this hampers the use of the IRX--$\beta$ dust attenuation relation as a proxy for dust types in galaxies.

\end{itemize}

Our results, especially the redshift dependent IRX--$\beta$ dust attenuation relation and the suggested degeneracy between stellar age, dust geometry and dust type, have clear consequences for the interpretation of observations of $z>1$ galaxies where IR information is lacking. We look forward to future observations testing the redshift dependent trend we found in our simulations and hopefully being able to discriminate between different dust types, ultimately testing the applicability of the IRX--$\beta$ dust attenuation relation in $z>0$ SFMS galaxies. 

\section*{Acknowledgements}
It is a pleasure to thank Martinna Donari and Vicente Rodriguez-Gomez for useful discussions. We thank the referee for their constructive comments. FM is supported by the Program "Rita Levi Montalcini" of the Italian MIUR. Simulations for this work were performed on the Draco and Isaac supercomputers at the Max Planck Computing and Data Facility. 

\section*{Data availability}
The data that support the findings of this study are available on request from the corresponding author. Most of the data pertaining to the IllustrisTNG project is in fact already openly available on the IllustrisTNG website, at \url{www.tng-project.org/data}; those of the TNG50 simulation, in particular, are expected to be made publicly available within some months from this publication, at the same IllustrisTNG repository.




\bibliographystyle{mnras}
\bibliography{lit} 




\appendix

\section{Checks on the Fiducial Choices and Fiducial Results} 
\label{Appendix}

\begin{table*}
    \centering
    \resizebox{\hsize}{!}{%
    \begin{tabular}{llll}
    \hline
    \textbf{{\sc skirt} setting} & \textbf{Our choice} & \textbf{Choice of \citealt{Vogelsberger2019}} & \textbf{Choice of \citealt{Gomez}}\\
    \hline
    \textit{Radiation sources} & & & \\
    \hline
    Old stars SED & {\sc galaxev} (Bruzual \& Charlot + Padova isochrones) & {\sc fsps} (MILES + MIST isochrones) & {\sc galaxev} (Bruzual \& Charlot + Padova isochrones) \\
    Young stars SED & {\sc mappings-iii} & {\sc mappings-iii} & {\sc mappings-iii}\\
    \hline
    \textit{Dust} \\
    \hline
    Dust grid & Imported Voronoi grid & Refined octree grid & Imported Voronoi grid\\
    Dust composition & Zubko MW dust mix & \citealt{Draine2007} MW dust mix & Zubko MW dust mix\\
    Selfabsorption & Turned off & Turned off & Turned off \\
    \hline
    Wavelength grid & 59 points in $\SI{0.1}{\micro\metre} < \lambda < \SI{1001}{\micro\metre}$ & 1168 points in $\SI{0.05}{\micro\metre} < \lambda < \SI{5}{\micro\metre}$ & 50 points in $\SI{0.35}{\micro\metre} < \lambda < \SI{0.95}{\micro\metre}$\\
    Photon package number & $10^6$ & equal to the total number of bound stellar particles, with $10^2 < N_\text{p} < 10^5$ & 50 per data cube pixel\\
    \hline
    \textit{Observing instrument}     & & & \\
    \hline
    Viewing distance & 10 Mpc & 1M pc & 10 Mpc\\
    Field of view & $15 r_{1/2}$ & variable & $15 r_{1/2}$\\
    Viewing angle & Along z-axis & Along z-axis & Along z-axis\\
    Output & SED and $100\times 100$ data cube & SED and data cube & SED and data cube with pixel scale $0.25 \text{arcsec}/\text{pixel}$ \\
    \hline
    \end{tabular}
    }
    \caption{A summary of the {\sc skirt} settings we chose, compared to other works based on the IllustrisTNG simulation suite. For an explanation of the self--absorption setting, see Appendix \ref{AppSelf}.}
    \label{tab:SKIRTset}
\end{table*}

\begin{table*}
  \centering
  \resizebox{\hsize}{!}{%
	\begin{tabular}{lllll} 
	    \hline
		\textbf{Parameter} & \textbf{Description} & \textbf{Origin (this work: Schulz et al. 2019)} & \textbf{Origin (\citealt{Vogelsberger2019})} & \textbf{Origin (\citealt{Gomez})}\\
		\hline
		  & \textit{Old Stellar Populations (age > 10 Myr)} & & & \\
		 \hline
		$h$ & smoothing scale & calculated as 32-th neighbour distance & calc. as 64th n. dist. & calc. as 32th n. dist.\\
        $M_\text{init}$ & initial mass of the stellar particle & TNG50 & TNG50 & TNG100\\
        $Z$ & metallicity of the stellar particle & TNG50 & TNG50 & TNG100 \\
        $t$ & age of the stellar particle & TNG50 & TNG50 & TNG100\\
		\hline
		& \textit{Young Stellar Populations (age < 10 Myr)} & & &\\
		\hline
		$h$ & smoothing scale & calculated as 32-th neighour distance & calc. as 64th n. dist. & calc. as 32th n. dist.\\
		SFR & star formation rate of the HII-region & calculated from initial mass ($M_\text{init}$) given by TNG50 & calc. from $M_\text{init}$ of TNG50 & calc. from $M_\text{init}$ of TNG100\\
		$Z$ & metallicity of the HII-region & TNG50 & TNG50 & TNG100\\
		$C$ & compactness parameter & free parameter, log$C = 5$ & free parameter, log$C = 5$ & free parameter, log$C = 5$\\
		$P_\text{ISM}$ & pressure of the ambient ISM & free parameter $\log_{10} \frac{P_\text{ISM}/k_B}{\SI{}{\centi \metre^{-3}}\SI{}{\kelvin}} = 5$ & free parameter, $\log_{10} \frac{P_\text{ISM}/k_B}{\SI{}{\centi \metre^{-3}}\SI{}{\kelvin}} = 5$ & free parameter, $\log_{10} \frac{P_\text{ISM}/k_B}{\SI{}{\centi \metre^{-3}}\SI{}{\kelvin}} = 5$\\
		$f_\text{PDR}$ & dust covering fraction of the PDR region & free parameter, $f_\text{PDR} = 0.2$ & free parameter, $f_\text{PDR} = 0.2$ & free parameter, $f_\text{PDR} = 0.2$\\
		\hline
		& \textit{Dust Distribution} & & &\\
		\hline
		$\rho_\text{gas}$ & gas cell density & TNG50 & TNG50 & TNG100 \\
		$Z$ & gas cell metallicity & TNG50 & TNG50 & TNG100 \\
		$f$ & fraction of metals locked up in dust & free parameter, $f = 0.3$ & calibrated, $0.24 \leq f(z) \leq 0.90$ for $4 \geq z \geq 2$ & free, $f = 0.3$\\
		$T_\text{max}$ & highest temperature at which gas contains dust & free parameter, $T_\text{max}=$ 75000 K & free parameter, $T_\text{max}=8000$ K & no $T_\text{max}$\\
		\hline
		
	\end{tabular}
	}
	\caption{Input parameters for {\sc skirt}, and their origin in our work, in \protect\citealt{Vogelsberger2019} and in \protect\citealt{Gomez}. Some values are taken directly from TNG50, some values are calculated out of TNG50--related data, others are chosen as free parameters. In \protect\citealt{Vogelsberger2019}, the dust--to--metal fraction was calibrated by comparing with observations.}
	\label{tab:SKIRT}
\end{table*}

In this Appendix, we quantify how different choices on the underlying radiative transfer and dust models may affect our results. Tables~\ref{tab:SKIRTset} and \ref{tab:SKIRT} summarize our {\sc skirt} fiducial setup and the adopted {\sc skirt} input parameters in our fiducial model. 
Here, we comment upon or explicitly show the effects of changing selected and relevant choices from tests performed on a subset of 440 randomly selected galaxies at $z = 0.5$. 
We distinguish between physical and {\sc skirt} implementation choices and comment on the possible effects of numerical resolution.

\subsection{Physical choices}

\subsubsection{Varying the Gas Dust Fraction}\label{Varying Fraction}
\label{app:fdtm}

\begin{figure*}
    \centering
    \includegraphics[width = \hsize]{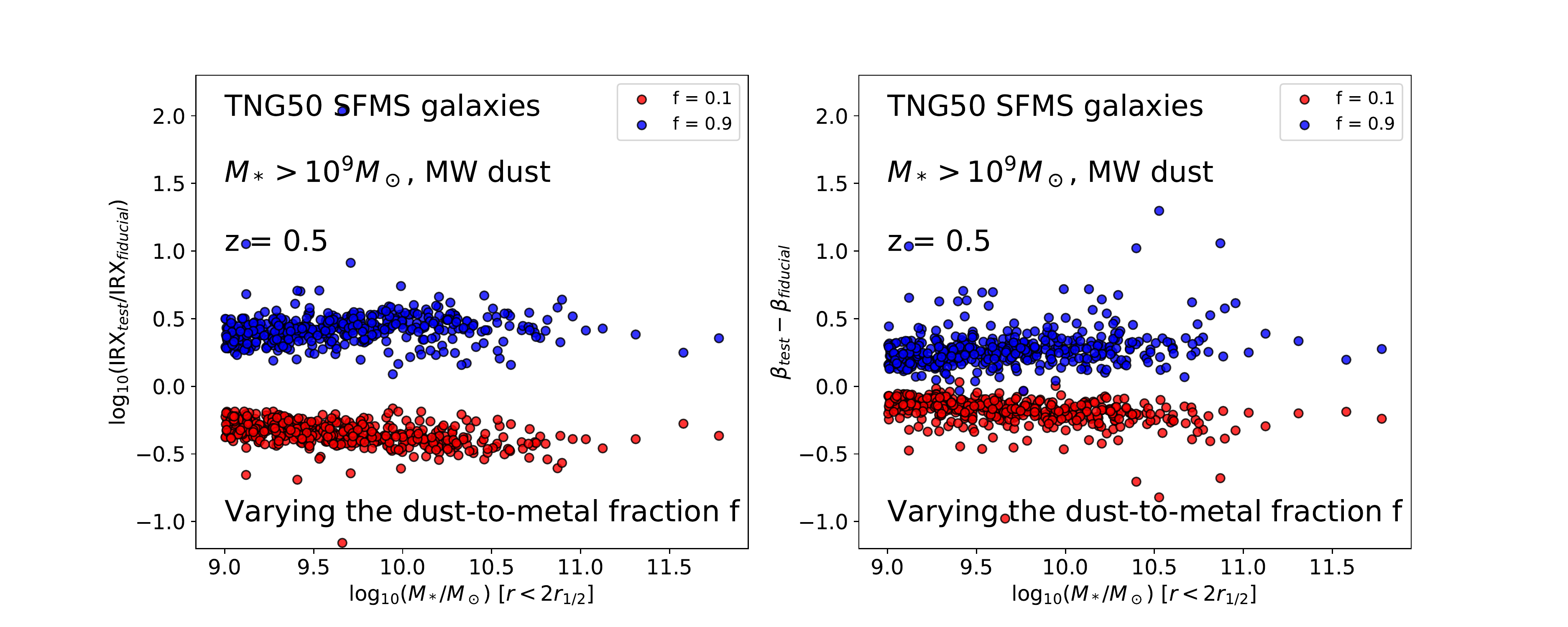}
    \includegraphics[width = \hsize]{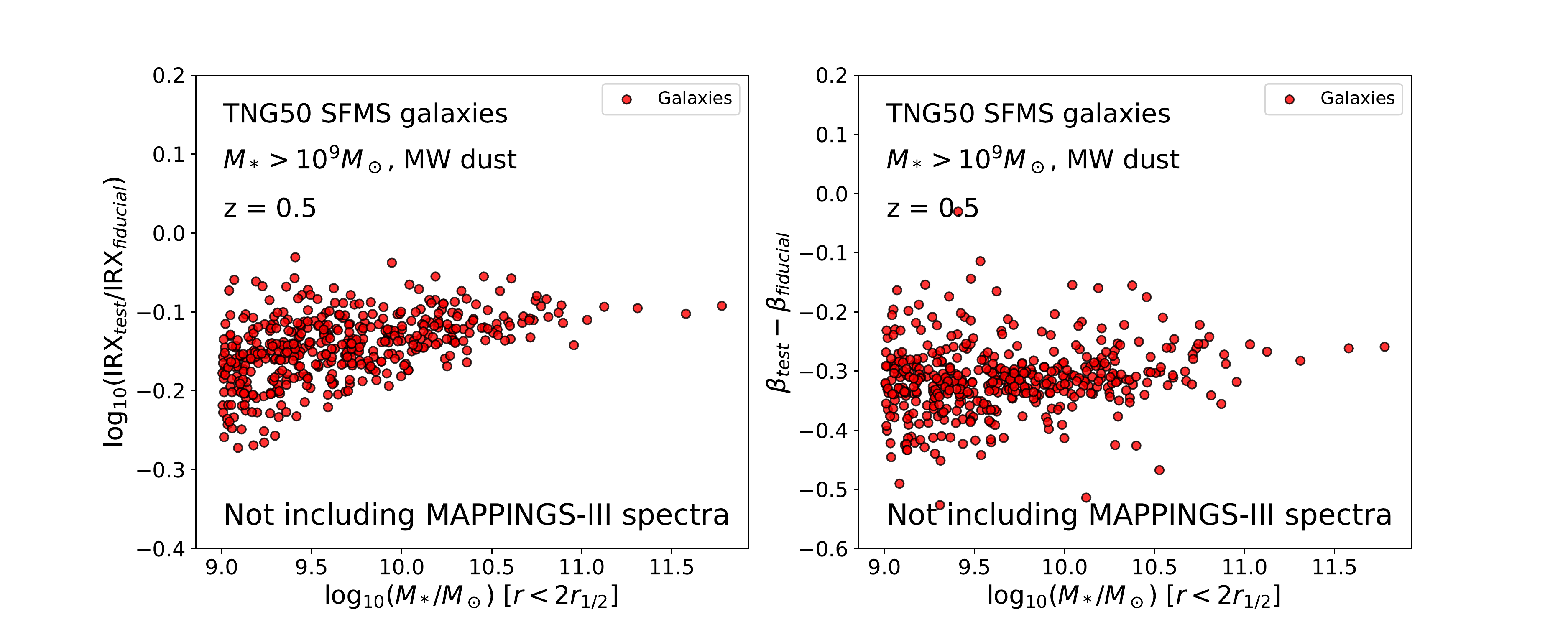}
    \includegraphics[width = \hsize]{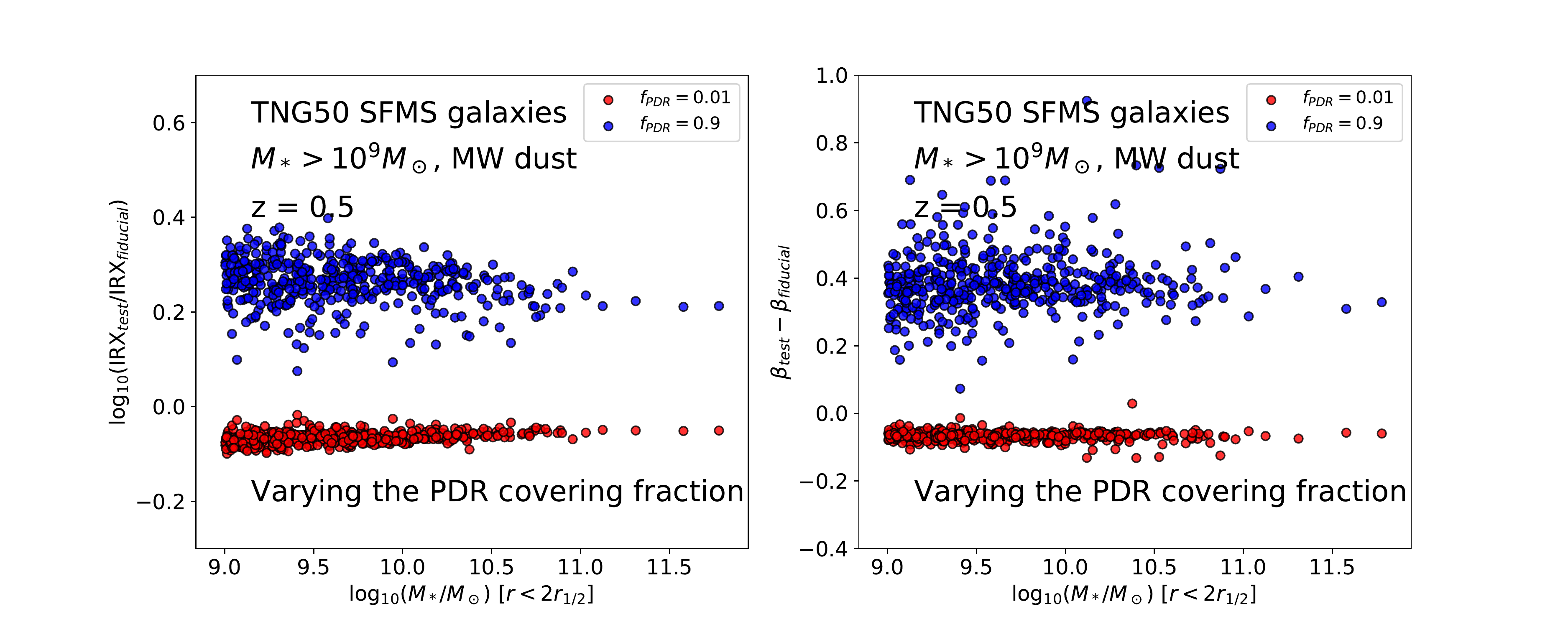}
    \caption{ Tests on selected post processing choices on the inferred values of IRX (left) and $\beta$ (right) for a random subset of 440 galaxies at $z = 0.5$. In all panels, we show the shift between the test variations and the fiducial choice, and here we show only the choices that we have found to have a non negligible impact in the quantitative results. Top: Effects of varying the dust-to-metal ratio from the fiducial value of 0.2 to 0.1 (red) and 0.9 (blue data points). Middle: Effects of excluding the {\sc mappings-iii} spectra for young stellar populations in the radiative transfer model from the fiducial setup. Bottom: Within the {\sc mappings-iii} implementation, effects of varying the dust covering fraction of the PDR regions from 0.2 (fiducial choice) to 0.01 (red) or 0.9 (blue data points). }
    \label{fig:fig15}
\end{figure*}

\begin{figure}
    \centering
    \includegraphics[width = \hsize]{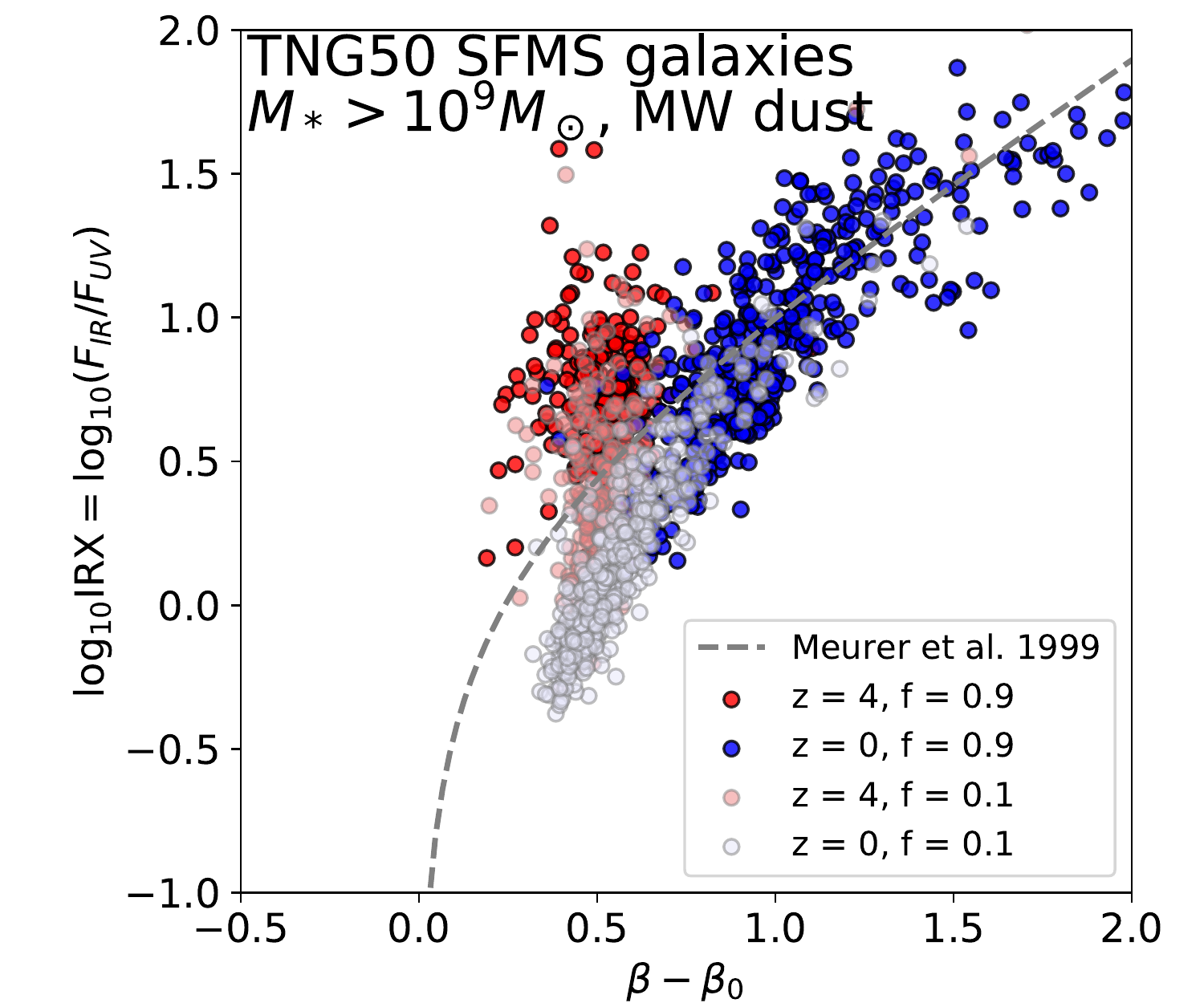}
    \caption{The IRX-($\beta-\beta_0$) distribution of $z=0$ and $z=4$ galaxies at different dust--to--metal ratios $f=0.1$ and $f=0.9$. While the $z=0$ population remains close to the local reference relation of Meurer when varying $f$, the $z=4$ galaxies follow a steeper curve, increasing their IRX while $\beta-\beta_0$ stays constant.}
    \label{fig:edactest}
\end{figure}

In the fiducial model, we set the dust--to--metal fraction of the gas to be $f = 0.3$ (see Eq. \ref{eq:eq2}). \citealt{Vogelsberger2019} performed a calibration with the TNG simulations to infer $f$ at each redshift via comparison to observed UV luminosity functions. They find f to vary between about 0.9 at $z\sim0$ to 0.1 at $z\sim6$. In the following, we test what effect a change of this parameter has on the IRX--$\beta$ distribution of the TNG50 SFMS galaxies. 
In principle, higher values of $f$ imply increased gas cell opacity in each galaxy, because each cell contains more dust. For a homogeneous dust screen, a higher opacity would lead simply to a shift along the respective local IRX--$\beta$ reference relations, toward higher IRX, and higher $\beta$. Only if the dust geometry is very complex (dust and young stellar populations are not cospatial) would a change in $f$ result in a more complex track of galaxies in the IRX--$\beta$ plane (e.g., see the explanation in \citealt{Casey_2014} and \citealt{Popping}). 
We vary the dust--to--metal fraction for a random subset of 440 galaxies at $z = 0.5$ in our sample and see how this affects their IRX and $\beta$ values (Figure~\ref{fig:fig15}, top row).  The difference between $f = 0.3$ and $f = 0.1$ (red symbols) is obviously smaller than between $f = 0.3$ and $f = 0.9$ (blue symbols): as the dust--to--metal fraction increases, log$_{10}$ IRX and $\beta$ increase as well, with an upward shift of about 0.5 and 0.3 on average, respectively. Still, we verified that the TNG50 galaxies at $z = 0.5$ are still roughly consistent with the $z\sim0$ observed IRX--$\beta$ dust attenuation relations.

Figure \ref{fig:edactest} highlights the different behaviour of $z=4$ galaxies with respect to $z=0$ galaxies when varying the dust--to--metal fraction $f$ (whis effectively varies their ISM dust opacity), after accounting for the intrinsic UV--slope. While at $z=0$ the galaxies broadly agree with the observed local reference relation by Meurer (increasing their opacity moves them along the relation to higher IRX and $\beta-\beta_0$), at $z=4$ galaxies trace a steeper track in the IRX--($\beta$-$\beta_0$) plane, implying they have a higher FUV--attenuation. Taking into account that the dust composition is fixed to Milky Way dust in our model, this supports the idea that at $z=4$ the geometry of the ISM dust distribution is less homogeneous than at low redshifts, changing their EDACs accordingly.

\subsubsection{Self absorption}\label{AppSelf}

During the {\sc skirt} setup, it is possible to turn on dust self absorption. When this option is turned on, the Monte-Carlo simulation will take into account absorption and re-emission of dust that has previously been re-emitted by dust itself. This effect only plays a role where very thick layers of dust are present, but significantly increases computation time. For this reason, in our fiducial model we have left this option turned off.  The test with a subset of $z=0.5$ galaxies (not shown) confirms the bounty of our choice, as with or without dust self absorption results on log$_{10}$ IRX and $\beta$ are uncertain to the percent level and below. Only for a fraction of massive galaxies the value of  log$_{10}$ IRX can differ by up to 5 per cent or so.

\subsubsection{{\sc mappings-iii} Parameters}\label{AppMappings}

In our fiducial setup, young stellar populations of ages younger than 10 Myr are assigned {\sc mappings-iii} spectra. A number of choices are needed to implement this, but they are overall all less relevant than the effect of using {\sc mappings-iii} spectra itself.

In Figure~\ref{fig:fig15}, middle row, we show the effects of turning off {\sc mappings-iii} completely and of only assigning Bruzual--Charlot spectra to the stellar particles regardless of their age. The impact is not negligible, at least at $z=0.5$ as shown. If we exclude {\sc mappings-iii} from our radiative transfer calculations, we end up with on average lower IRX (a shift of about 0.1-0.2 in the log) and lower $\beta$ (an average shift of 0.3-0.35), because the effects of birth clouds are neglected. The shift in IRX is more relevant (up to 0.25 dex) for lower-mass galaxies. 
We keep {\sc mappings-iii} spectra included in our fiducial model, to be more realistic and comparable to other works that employed {\sc skirt} radiative transfer calculations to IllustrisTNG.

Among the physical choices needed to implement the  {\sc mappings-iii} emission, the interstellar medium pressure has no significant effect on the results (smaller than per cent level shifts: not shown). 
The compactness parameter $C$ is directly influenced by $P_\text{ISM}$ via the equation:

\begin{equation}
    \log_{10} C = \frac{3}{5} \log_{10} \frac{M_{cl}}{\MSUN}+\frac{2}{5}\log_{10} \frac{P_\text{ISM}/k_B}{\SI{}{\centi \metre^{-3}}\SI{}{\kelvin}}
\end{equation}

where $M_{cl}$ is the  star cluster mass. As explained in section \ref{Methods}, stellar particles have masses of around $10^5 \MSUN$ in the highest resolution variant of TNG50. Therefore the compactness derives itself directly from the interstellar medium pressure via:

\begin{equation}
    \log_{10} C = 3+\frac{2}{5}\log_{10} \frac{P_\text{ISM}/k_B}{\SI{}{\centi \metre^{-3}}\SI{}{\kelvin}}
\end{equation}

 However, these choices are not impactful in any manner that is relevant for the study at hand. We therefore choose as fiducial the value specified in the literature of $X = \log_{10} \frac{P_\text{ISM}/k_B}{\SI{}{\centi \metre^{-3}}\SI{}{\kelvin}} = 5$ for our fiducial model.

On the other hand, an increase of the PDR covering fraction to $f_\text{PDR} =0.9$ would not be negligible. In Figure~\ref{fig:fig15} this effect is quantified by comparing the fiducial choice of $f_\text{PDR} = 0.2$ with $f_\text{PDR} = 0.01$ (red symbols) and $0.9$ (blue symbols) for a subset of test galaxies at $z=0.5$. 
A higher covering fraction of birth clouds would increase the absorption of UV--emission by the young stellar populations (effectively increasing $\beta$) and in turn increase the infrared emission. This only happens, however, when we choose extreme values such as $f_\text{PDR} = 0.9$, and therefore we maintain that the qualitative analysis of our results is not affected by this.

\subsection{{\sc skirt} implementation choices}

{\sc skirt} offers various choices for e.g. the selection of the wavelength grid which is used for sampling the spectra and the number of photon packages for the radiative transfer.

Throughout, we chose a custom wavelength grid with 59 wavelengths, with high detail in the UV and IR ranges. To test if the results given by this choice are converged, we increase the number of wavelengths to 93 wavelengths and see if the IRX and $\beta$ results change. Differences are minimal across the mass range (less than a few percent for log$_{10}$ IRX and less than 10 percent for $\beta$, respectively), giving us confidence that the IRX--$\beta$ results are converged already with 59 wavelengths.

To test if our choice of number of photon packages grants convergence, we increase the photon package count tenfold, and rerun the radiative transfer on a subset of 267 galaxies at $z = 0.5$. Differences are smaller than at the per cent level, implying that our fiducial choice is converged.

Finally, in order to perform the {\sc skirt} simulation, we need to calculate adaptive smoothing scales for the stellar particles. In our fiducial setup we have chosen the smoothing scale of each particle to be the distance to the 32nd closest neighboring stellar particle. We tested if the results change when opting for smoothing scales that correspond to the distance to the 16th and 64th closest neighbors. We find no noticeable effects on the derived IRX and $\beta$ values (at per cent levels).

\begin{figure*}
    \centering
    \includegraphics[width = \hsize]{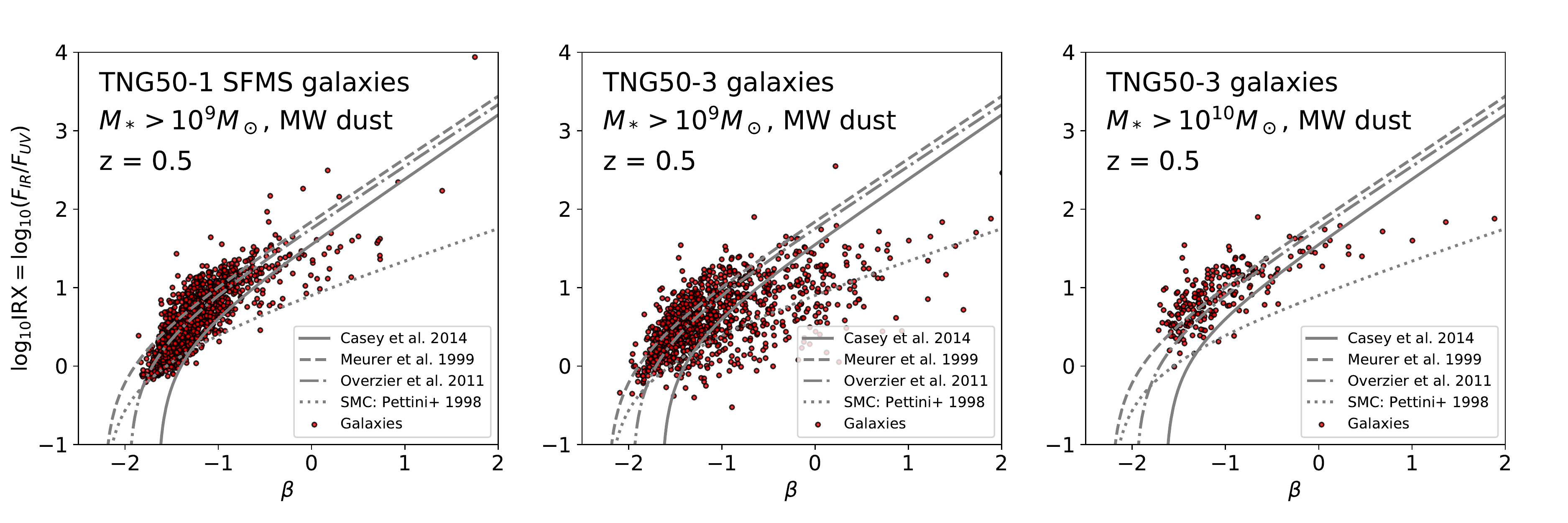}
    \caption{ Effects of numerical resolution on the resulting IRX--$\beta$ distribution at $z = 0.5$. Left: results from our flagship run, TNG50 aka TNG50-1. Mid and right: IRX--$\beta$ distributions from the low-resolution variant TNG50-3, with 64 times worse mass resolution, for galaxies with mass $M_* > 10^{9}$ and $M_* > 10^{10}\MSUN$, respectively.}
    \label{fig:app5}
\end{figure*}

\begin{table}
  \centering
  \resizebox{\columnwidth}{!}{%
	\begin{tabular}{cccccccc} 
		\hline
		Run Name & $L^{z=0}_\text{box}$ & $N_\text{gas}$ & $N_\text{DM}$ & $m_\text{baryon}$ & $m_\text{DM}$ & $\epsilon^{z=0}_\text{DM, stars}$ & $\epsilon^\text{min}_\text{gas}$\\
		 & [$Mpc$] & & & [$10^6 M_\odot$] & [$10^6 M_\odot$] & [kpc] & [kpc]\\
		\hline
		TNG50-1 & 51.7 & $2160^3$ & $2160^3$ & 0.085 & 0.45 & 0.288 & 0.074 \\
        TNG50-3 & 51.7 & $540^3$ & $540^3$ & 5.42 & 29.04 & 1.152 & 0.295\\
		\hline
	\end{tabular}
	}
	\caption{A summary of the most important numerical specifications of the highest resolution TNG50 variant (TNG50-1), which is used in the main body of this work, and of a low resolution TNG50 variant used in the resolution test in Appendix \ref{AppResolution} (TNG50-3). $L^{z=0}_\text{box}$ denotes the proper length of the simulation box at $z=0$. $N_\text{gas}$ is the number of gas cells at the start of the simulation run, which can change over the course of the simulation due to star formation or refinement of grid cells. $N_\text{DM}$ is the number of dark matter particles, which stays constant over the whole simulation. $m_\text{baryon}$ is the mean gas cell mass, and also roughly the initial mass of stellar particles, while $m_\text{DM}$ is the mass of the dark matter particles (all DM--particles have the same mass). The gravitational softening length for dark matter particles at $z=0$ is given by $\epsilon^{z=0}_\text{DM, stars}$. Softening lengths for gas cells are adaptive, with a minimum value of $\epsilon^\text{min}_\text{gas}$ (see e.g. Fig.1 of \citealt{Pillepich50}). }
	\label{tab:numerics}
\end{table}

\subsection{Resolution Effects}\label{AppResolution}

 To uncover possible effects of numerical resolution, we qualitatively compare the IRX--$\beta$ distributions of our flagship simulation, TNG50, with one with much worse resolution. For this purposes, we use a low--resolution variant of TNG50, namely TNG50-3 (see Table \ref{tab:numerics}): this has only $2\times540$ resolution elements inside the same comoving volume. The mass resolution of the baryonic resolution elements is therefore $54\times 10^5 \MSUN$ and for dark matter resolution elements it is $290\times 10^5 \MSUN$: 64 times worse than TNG50. In Figure~\ref{fig:app5}, we compare the IRX--$\beta$ distribution of all selected $z = 0.5$ galaxies of the high resolution variant TNG50 (aka TNG50-1) with the low resolution variant TNG50-3. A large number of galaxies in TNG50-3 fall below the reference relations. This is due to the fact that galaxies with masses of $10^9 \MSUN$ are not well resolved, as they only consist of a few hundreds of stellar particles in total. On the right hand side of the figure, we therefore experiment by excluding TNG50-3 galaxies with stellar masses below $10^{10} \MSUN$, so as to only include reasonably resolved galaxies. When including only galaxies of TNG50-3 that are similarly resolved as in TNG50 (with a few thousand stellar particles each), the reference relation is recovered, as in TNG50. This underlines the importance of using well resolved galaxies (both in mass and spatial numerical resolution) but, conversely, also shows that within TNG50 and above the adopted minimum mass of $10^{9} \MSUN$ (i.e. more than $10^4$ stellar particles each) the results provided throughout are trustworthy.

\bsp	
\label{lastpage}
\end{document}